\documentclass[a4paper]{amsart}
\usepackage{latexsym}
\usepackage{amsmath}
\usepackage{amssymb}
\usepackage{amsfonts}
\usepackage{hyperref}
\usepackage{color}
\usepackage{enumitem}
\definecolor{Blue}{rgb}{0.,0.,1.}
\definecolor{Red}{rgb}{1.,0.,0.}

\newcounter{smallarabics}
\newenvironment{arabicenumerate}
{\begin{list}{{\normalfont\textrm{(\arabic{smallarabics})}}}
  {\usecounter{smallarabics}\setlength{\itemindent}{0cm}
   \setlength{\leftmargin}{5ex}\setlength{\labelwidth}{4ex}
   \setlength{\topsep}{0.75\parsep}\setlength{\partopsep}{0ex}
   \setlength{\itemsep}{0ex}}}
{\end{list}}

\newcounter{smallroman}

\newcommand{\ben}{\begin{arabicenumerate}}  
\newcommand{\een}{\end{arabicenumerate}}

\def\init{\setcounter{equation}{0}}

\newtheorem{theoreme}{Theorem }[section]
\newtheorem{proposition}[theoreme]{Proposition}
\newtheorem{lemma}[theoreme]{Lemma}
\newtheorem{definition}[theoreme]{Definition}
\newtheorem{corollary}[theoreme]{Corollary}
\newtheorem{remark}[theoreme]{Remark}
\newtheorem{example}[theoreme]{Example}
\def\beq#1\eeq{\begin{align}#1\end{align}}
\def\beqa#1\eeqa{\begin{align}#1\end{align}}
\def\bes#1\ees{\begin{split}#1 \end{split}}

\newcommand{\DD}{D}

\newcommand{\bex}{\begin{example}}
\newcommand{\eex}{\end{example}}
\def\bel{\begin{lemma}}
\def\eel{\end{lemma}}
\def\bet{\begin{theoreme}}
\def\eet{\end{theoreme}}
\def\bed{\begin{definition}}
\def\eed{\end{definition}}
\def\ber{\begin{remark}}
\def\eer{\end{remark}}

\def\rr{{\mathbb R}}

\def\cc{{\mathbb C}}
\def\nn{{\mathbb N}}

\def\slim{{\rm s-}\lim}
\def\cl{{\rm cl}}
\def\H{{\rm H}}

\def\al{\alpha}
\def\i{{\rm i}}
\def\Span{{\rm Span}}
\def\qed{$\Box$\medskip}
\def\Sp{{\mathcal Sp}}
\def\D{\mathcal D}
\def\Ran{\mathrm{Ran}}
\def\bbbone{{\mathchoice {\rm 1\mskip-4mu l} {\rm 1\mskip-4mu l}
{\rm 1\mskip-4.5mu l} {\rm 1\mskip-5mu l}}}
\def\one{\bbbone}
\def\cH{{\mathcal H}}
\def\C{Q}
\def\R{{\langle R\rangle}}

\def \p{ \partial}

\def\12{\frac{1}{2}}

\def\xt{\frac{x}{t}}
\def\supp{{\rm supp}}

\def\e{{\mathrm e}}

\def\bep{\begin{proposition}}
\def\eep{\end{proposition}}
\newcommand{\n}{\ell}
\newcommand{\tiball}{\mathcal{\ti{B}}}
\newcommand{\ball}{\mathcal B}
\newcommand{\Theorem}{Thm. }
\newcommand{\Theorems}{Thms.}
\newcommand{\Proposition}{Prop.}

\newcommand{\Definition}{Def.}
\newcommand{\pairs}{particles }
\newcommand{\cpf}{\mathrm{cpf}}

\newcommand{\hilnp}{\hil^{\mathrm{cpf} } }

\newcommand{\hy}{\un y}
\newcommand{\hxi}{\un \xi}
\newcommand{\kh}{h}

\newcommand{\pha}{\phantom}

\newcommand{\ti}{\tilde}
\newcommand{\w}{\mathrm{w}}
\newcommand{\un}{\underline}

\newcommand{\De}{\Delta}

\newcommand{\mcL}{\mathcal{L}}

\newcommand{\lan}{\langle}
\newcommand{\ran}{\rangle}

\newcommand{\pa}{\partial}

\newcommand{\Om}{\Omega}
\newcommand{\si}{\sigma}

\newcommand{\de}{\delta}

\newcommand{\Ga}{\Gamma}

\newcommand{\hil}{\mathcal{H}}
\newcommand{\om}{\omega}
\newcommand{\mfa}{\mathfrak{A}}
\newcommand{\mco}{\mathcal{O}}

\newcommand{\eps}{\varepsilon}
\newcommand{\fr}[2]{\frac{#1}{#2}}

\newcommand{\real}{\mathbb{R}}

\newcommand{\la}{\lambda}
\newcommand{\ov}{\overline}

\newcommand{\non}{\nonumber}
\newcommand{\complex}{\mathbb{C}}

\newcommand{\hx}{\un x}
\def\tom{\tilde{\omega}}

\newcommand{\setbar}{\ : \ }
\newcommand{\BB}{\un B}      %{B_1,B_2}
\newcommand{\BBc}{\ov B}
\newcommand{\wh}{\widehat}
\newcommand{\timeso}{\overset{\mathrm{out}}{\times}}

\def\rx{x}
\def\xt{\frac{x}{t}}

\def\rr{\real}
\def\cH{\hil}
\def\bbbone{{\mathchoice {\rm 1\mskip-4mu l} {\rm 1\mskip-4mu l}
{\rm 1\mskip-4.5mu l} {\rm 1\mskip-5mu l}}}
\def\one{\bbbone}
\def\cS{\mathcal{S}}
\def\supp{{\rm supp}}

\def\p{\partial}
\def\nn{{\mathbb{N}}}
\def\12{\frac{1}{2}}
\def\Dom{{\rm Dom}\ }

\def\coinf{C_{0}^{\infty}}
\def\cc{{\mathbb{C}}}
\def\tom{\tilde{\omega}}

\usepackage{verbatim}

\DeclareFontFamily{U}{mathx}{\hyphenchar\font45}
\DeclareFontShape{U}{mathx}{m}{n}{
      <5> <6> <7> <8> <9> <10>
      <10.95> <12> <14.4> <17.28> <20.74> <24.88>
      mathx10
      }{}
\DeclareSymbolFont{mathx}{U}{mathx}{m}{n}
\DeclareFontSubstitution{U}{mathx}{m}{n}
\DeclareMathAccent{\widecheck}{0}{mathx}{"71}
\DeclareMathAccent{\wideparen}{0}{mathx}{"75}

\newlength{\dinwidth}
\newlength{\dinmargin}
\begin{document}

\title{A criterion for asymptotic completeness in local relativistic QFT}
\author{Wojciech Dybalski}
\address{ Institut f\"ur Theoretische Physik,  ETH Z\"urich, 8093 Z\"urich
Switzerland 
and 
Zentrum Mathematik, Technische Universit\"at M\"unchen,
D-85747 Garching Germany}
\email{dybalski@ma.tum.de}
\author{Christian G\'erard}
\address{D\'epartement de Math\'ematiques, Universit\'e de Paris XI, 91405 Orsay Cedex France}
\email{christian.gerard@math.u-psud.fr}
\keywords{local quantum field theory, Haag-Ruelle scattering theory, Araki-Haag detectors, asymptotic completeness}
\subjclass[1991]{81T05, 81U99}
\begin{abstract}

We formulate a generalized concept of asymptotic completeness 
and show that it holds in any Haag-Kastler quantum field theory with an upper and lower mass gap. 
It  remains valid  in the presence of pairs of oppositely charged particles in the vacuum sector, which invalidate the conventional property of asymptotic completeness. Our result can  be restated as a criterion characterizing a class of theories with  complete particle interpretation in the conventional sense. This criterion is formulated in terms of certain asymptotic observables (Araki-Haag detectors) whose existence, as strong limits of their approximating sequences, is our main technical result. It is proven with the help of a novel propagation estimate, which is also relevant to scattering theory of quantum mechanical dispersive systems.

\end{abstract}
\maketitle

\section{Introduction}\label{Intro}
\setcounter{equation}{0}

The physical interpretation of local relativistic quantum field theories (QFT) in terms of particles is a long-standing open problem. 
The only known class of non-trivial  asymptotically complete models are the recently constructed two-dimensional 
theories with factorizing  $S$-matrices \cite{Le08, Ta13}. In the thoroughly studied $P(\phi)_2$ models only partial results on 
asymptotic completeness~(AC) of two- and three-particle scattering have been found \cite{SZ76,CD82}. The progress on this fundamental problem is hindered by several  conceptual and technical difficulties:  

(1) On the conceptual side we face a difficulty which is typical for QFT:  the algebra of observables 
of a system with infinitely many degrees of freedom may have many non-equivalent representations (`sectors') labelled by some `charge' \cite{DHR71,DHR74,BF82}.  Thus the vacuum
sector, whose Hilbert space $\hil$ consists of states of zero charge, may contain collections of charged particles whose total charge is zero, for example pairs of oppositely charged excitations.
As such configurations do not belong to the subspace $\hil^+$ of Haag-Ruelle scattering states of neutral particles,
they undermine the conventional AC relation:
\beq
\hil^+=\hil, \label{conventional-AC}
\eeq
inherited from quantum mechanics. 

(2) Even if a theory has trivial superselection structure, or all its superselection sectors are properly taken into account, the conventional AC may fail due to the presence of (unphysical) states with too many local degrees of freedom, 
which do not admit any particle interpretation. This is the case in certain generalized free fields  \cite{Gre61, HS65}.

(3) On the technical side the main stumbling block  is our poor understanding   of  dynamics
of dispersive systems i.e., systems of particles with non-quadratic dispersion relations. We recall in 
this context that the classical results on the 
$n$-body AC in quantum mechanics \cite{SiSo87,Gr90, De93} do not apply to such theories.

In essence, the first two problems above mean that it is not possible to prove conventional  AC  
from the Haag-Kastler postulates, since there exist counterexamples of physical (1) and unphysical (2) type. It is at best possible to 
formulate criteria which characterize a class of theories  for which (\ref{conventional-AC}) holds. A  search for such conditions, 
initiated almost half a century ago in \cite{HS65} and continued in  \cite{Bu87, Bu94, BS05}, has so far been unsuccessful. 
In the present work we formulate a model-independent criterion for conventional AC in massive Haag-Kastler QFT.
Our analysis can be summarized as follows:  To tackle difficulty (1), we introduce a `charged \pairs free' subspace $\hilnp\subset\hil$, defined in (\ref{charged-particles-free}) below. This subspace is constructed with the help of suitable asymptotic observables (generalizations of the Araki-Haag detectors \cite{AH67}), sensitive only to neutral particles. 
We formulate a generalized (weaker) concept of AC, suitable  for theories with non-trivial superselection structure, which requires that  
\beq
\hil^+=\hilnp.  \label{main-result}
\eeq
We show that this variant of AC holds in any Haag-Kastler  QFT with an upper and lower mass gap, as defined in 
 Subsection~\ref{general-assumptions} below. This class includes non-trivial models, as for example $\la \phi^4_2$ and
$\la \phi^4_3$ theories at small $\la$ \cite{GJS73, Bur77}.
 Incidentally,  relation~(\ref{main-result}) shows that also the unphysical states of type (2) are 
eliminated from the `charged \pairs free' subspace.   Equality (\ref{main-result}) can   immediately be reformulated
as a criterion for conventional AC:
\beq
\hil=\hil^{\cpf}\quad  \Leftrightarrow \quad   \hil^+=\hil. 
\label{criterion-for-AC}
\eeq

Our proof of relations~(\ref{main-result}), (\ref{criterion-for-AC}) relies on deep similarities between non-relativistic and 
relativistic scattering theory  brought to light in our recent work \cite{DG12}. 
They allow us to apply powerful quantum-mechanical  techniques, as for example the method of propagation estimates \cite{SiSo87},
in the relativistic setting. At this technical level we encounter  difficulty (3): The approach of Graf \cite{Gr90},  
which relies on a phase space propagation estimate,  does not apply in the presence of three or more particles with relativistic dispersion relations.  
We solve this problem with the help of a novel propagation estimate (\Proposition~\ref{novel-propagation}) which is the main  
technical result of this work. We expect that it will also find applications in scattering theory of non-relativistic dispersive systems \cite{Zi97,Ge91}.

The question if criterion~(\ref{criterion-for-AC}) is useful for proving conventional AC in concrete interacting models is left open in the present work. Nevertheless, let us provide several remarks on this point which may indicate directions of future research: 
For theories with trivial superselection structure we expect that our criterion is sharp in the sense that it only eliminates 
unphysical examples of type (2). We recall in this context that general conditions for the absence of Doplicher-Haag-Roberts (DHR) 
sectors  in two-dimensional massive theories were given in \cite{Mu98}.  These conditions (Haag duality for double cones and split property for wedges) 
offer a more specific framework for future investigations of the problem of AC
in concrete interacting theories.
A class of examples which should fit into this setting are the $P(\phi)_2$ models in the one-phase region\footnote{Split property for wedges is 
expected but not known yet in these theories. Cf. Section~7 of \cite{Mu98}.}.

Theories with non-trivial superselection structure should be embedded into  larger theories, which take all the superselection
sectors into account, before  criterion (\ref{criterion-for-AC}) is checked. Such an embedding can, in principle, be 
accomplished for any massive  Haag-Kastler QFT by a suitable variant of the DHR construction  \cite{DHR71,DHR74}.
In particular, for massive theories in physical spacetime this procedure is very well understood \cite{BF82} and allows for a 
construction of Haag-Ruelle scattering states involving both neutral and charged particles. 
We recall, however, that the resulting larger theory contains charge carrying fields whose commutation and localization properties may significantly
differ from the familiar properties of observables: In physical spacetime 
they may have Fermi statistics and/or string-like localization. In spacetimes
of lower dimension braid group statistics \cite{FRS89} or soliton sectors \cite{Fr76,BFG78} may appear. 
The question of validity of  relations (\ref{main-result}), (\ref{criterion-for-AC}) in the presence of these interesting complications is left for future work. Examples of interacting theories with non-trivial superselection structure (soliton sectors) are the $P(\phi)_2$ models in the two-phase region\footnote{We refer to the Appendix of \cite{SW} and references therein for a discussion of superselection structure and its relation to the problem of  AC in $P(\phi)_2$ models.}.

To outline the construction of the `charged \pairs free' subspace $\hilnp$,
appearing in relations~(\ref{main-result}) and (\ref{criterion-for-AC}),
 we need some preparations. The restrictive form of the spectrum condition,
which we adopt in this work, is important for this discussion: We assume that
the spectrum of the energy-momentum operators, denoted $\Sp \, U$, consists of an isolated simple eigenvalue
at zero, corresponding to the vacuum vector $\Om$, an isolated mass hyperboloid $H_m:=\{(E,p)\in \real^{1+d}\,:\, E=\om(p)\}$,  $\om(p):=\sqrt{p^2+m^2}$, carrying neutral single-particle states of mass $m>0$ and a multiparticle spectrum $G_{2m}$ whose lower boundary is  $H_{2m}$. For precise definitions of other concepts appearing in the discussion below
the reader should consult Section~\ref{Framework}.

Let us fix an energy-momentum vector $\ti p=(E, p)\in H_m$,   and construct  time-dependent 
families of observables $t\mapsto C_t$ which are the main building blocks of $\hilnp$:
We choose an almost-local operator $B$ from the algebra of observables $\mfa$ of our theory,  s.t.  its energy-momentum transfer
belongs to a small neighbourhood of $-\ti p$. 
Denoting by $B(t,x)$  the translation of  $B$ by the spacetime vector $(t,x)$ and choosing a suitable function  on the phase space 
$h\in C_0^{\infty}( T^{*}\real^{d})$ we set   
\begin{align}
& C_{t}:= \int \, \kh^\w_{t}(x,y)    B^*(t,x)B(t,y)dxdy, 
\label{one-particle-detector}
\end{align}
where $h_{t}(x,\xi):=h(x/t,\xi)$, $h_{t}^\w\in B(L^{2}(\real^{d}))$ is the {\em Weyl quantization} of the symbol $h_{t}$ and 
$\kh^\w_{t}(x, y)$ is its integral kernel. The function $h$ essentially has the form $h(x,\xi)=h_{0}(x)\chi(x-\nabla\om(\xi))$, where  
$h_{0}\in C_0^{\infty}(\real^d)$ is supported in a small neighbourhood of the point $\nabla\om( p)$ and $\chi$ is supported in a 
small neighbourhood of zero.

Let us now justify that $t\mapsto C_{t}$  can be interpreted for large $t$ as a detector sensitive only to neutral particles whose energies-momenta belong to
a small neighbourhood of $\ti p$.  By computing the limit $C^+$ of  $t\mapsto C_t$, as $t\to\infty$  on the subspace of  Haag-Ruelle scattering states involving both neutral and charged particles, one obtains a counterpart of formula~(28) from \cite{AH67}:
\beqa
C^+=(2\pi)^d\sum_{q,q'}\int d\xi\, h\big(\nabla\om_q(\xi), \xi \big)
\lan \xi, q|B^*B| \xi,q'\ran  a_{+,q}^*(\xi) a_{+,q'}(\xi), \label{Araki-Haag-formula}
\eeqa
where  $\om_q(\xi):=\sqrt{\xi^2+m_q^2}$, $m_q$ is the mass of a particle of type $q$, $|\xi, q\ran$ its
plane-wave configuration with momentum $\xi$ and  $a_{+,q}^*(\xi)$ the asymptotic creation operator of such a configuration, given by the Haag-Ruelle theory. 
The sum in (\ref{Araki-Haag-formula}) extends over all pairs $q,q'$ s.t. $m_q=m_{q'}$.
In view of the relation $h(\nabla\om_q(\xi),\xi)=h_{0}(\nabla\om_q(\xi)  )\chi(\nabla\om_q(\xi)-\nabla\om(\xi))$ and
of the support properties of  $h_{0}$ and $\chi$, the function $\xi\mapsto h(\nabla\om_q(\xi),\xi)$ is non-zero
only for such $\xi$ that $(\om_q(\xi),\xi)$ is in a small neighbourhood of $\ti p$. (In particular, $m_q$ must be close to $m$).
 For such $\xi$ we also have 
\beqa
B|\xi, q\ran=|\Om\ran \lan \Om|B|\xi, q\ran, 
\eeqa
since the energy-momentum transfer of $B$ is close to $-\ti p$ and $m_q\geq m$ for all $q$ \footnote{
If the particle of type $q$ is neutral, we have $m_q=m$ since we assumed that there is only one isolated mass hyperboloid in $\Sp\, U$. 
If the particle of type $q$ is charged, we have $m_q\geq m$, since  otherwise the multiparticle spectrum $G_{2m}$ in the vacuum sector would start below $E=2m$ due to the presence of pairs of oppositely charged particles of mass $m_q$. }. 
If the particle of type $q$ is neutral, we can easily find $B$, within the above restrictions, s.t. $\lan\Om|B|q,\xi\ran\neq 0$.
However, if the particle of type $q$ is charged, we have $\lan\Om|B|q,\xi\ran=0$, since observables cannot create charged states from the
vacuum. Hence, the sum in (\ref{Araki-Haag-formula}) extends only over neutral particle types and the integral over such $\xi$ that $(\om(\xi),\xi)$
is in a small neighbourhood of $\ti p$. Thus any non-zero vector from the range of $C^+$, on the subspace of Haag-Ruelle scattering states,
contains a neutral  particle whose energy-momentum vector is in a small neighbourhood of $\ti p$ (and possibly some other neutral or charged particles).

We mention as an aside that for a symbol  $h(x, \xi)=h_{0}(x)$
we recover from (\ref{one-particle-detector}) a time-dependent family of observables of the form
\begin{align}
C_{t}^{\mathrm{AH}}:= \int h_{0}\left(\xt\right) B^{*}(t,x) B(t,x) dx \label{Araki-Haag-det}
\end{align}
which is the usual \emph{Araki-Haag detector} \cite{AH67}. Arguing as above 
one can justify that these detectors are sensitive only to particles whose velocities belong to the support of $h_0$ i.e.,
are in a neighbourhood of $\nabla\om(p)$. However,  one  cannot conclude in this case that the masses $m'$ of these particles are close to
$m$. Thus  $t\mapsto C_{t}^{\mathrm{AH}}$ is sensitive not only to neutral particles of mass $m$, but may also detect 
some neutral or charged particles whose mass hyperboloids are embedded in the multiparticle spectrum in the respective sector. 
(Charged particles with isolated mass hyperboloids can be excluded by exploiting the energy-momentum transfer of $B$, similarly
as above). While this sensitivity to other particles would disappear in the next step of our analysis, which concerns products
of detectors (see (\ref{n-detector-limit}) below), we find it conceptually more satisfactory to work from the outset 
with detectors~(\ref{one-particle-detector}), which are only sensitive to neutral particles of mass $m$. A more technical reason to 
use these detectors, related to difficulty (3),  will be discussed later on in this section.

Coming back to the construction of the `charged particles free'   subspace $\hil^{\cpf}$,  we  fix some open bounded set  
$\De\subset G_{2m}$, which is small compared to the mass gap, 
(i.e., s.t. $(\ov{\De}-\ov{\De})\cap \Sp \, U=\{0\}$) and let $\one_{\De}(U)$ be the corresponding spectral projection.  
We intend to characterize states from the range of $\one_{\De}(U)$ which  are configurations of $n\geq 2$ neutral particles of mass $m$.
Let us consider one such configuration consisting of particles whose energy-momentum vectors are centered around
some  $\ti p_i\in H_m$, $i=1,\ldots, n$, which satisfy  $\ti p_1+\cdots+\ti p_n\in \De$ and $\ti p_i\neq \ti p_j$ for $i\neq j$.
We denote by $t\mapsto C_{i,t}$, $i=1,\ldots,n$,  detectors of the form (\ref{one-particle-detector}) sensitive to neutral particles 
whose energy-momentum vectors  are close to $\ti p_i$. In particular, we require that the corresponding functions $h_i\in C_0^{\infty}( T^{*}\real^{d})$ have disjoint 
supports in the first variable.
A coincidence arrangement of this collection of detectors, defined as
\beqa
Q^+_n(\De)\Psi:=\slim_{t\to\infty}C_{1,t}\ldots C_{n,t}\Psi, \quad \Psi\in \one_{\De}(U)\hil, \label{n-detector-limit}
\eeqa
is an asymptotic observable sensitive to the prescribed configuration of $n$ neutral particles. In fact, for $\Psi$ from \emph{the subspace of
 Haag-Ruelle scattering states}, it follows from our discussion of individual detectors above that any vector from the range of 
$Q^+_n(\De)$ contains only the prescribed configuration
of neutral particles. (The presence of any  other  particles is energetically excluded,  since $Q^+_n(\De)$ commutes with   $\one_{\De}(U)$,
$\ti p_1+\cdots+\ti p_n\in \De$ and $\De$ is small compared to the mass gap). 
It is an important finding of the present paper  that the same holds for   \emph{any}
$\Psi\in \one_{\De}(U)\hil$, including the existence of the limit in (\ref{n-detector-limit}). 
Leaving the question of convergence in (\ref{n-detector-limit}) to the later part of this Introduction,  we set $\hil(\De)=\one_{\De}(U)\hil$ and define the $n$-particle component of the `charged \pairs free' subspace associated with the set $\De$ as 
\beqa
\hil^\cpf_{n}(\De):=\Span\{ Q^+_{n,\al}(\De)\hil(\De)\,:\, \al\in J\,\}^{\cl},
\eeqa
where the span extends over the collection of all the asymptotic observables of the form (\ref{n-detector-limit}), corresponding to various
configurations of $n$ neutral particles with total energy-momentum in $\De$. 
We show in \Theorems~\ref{Main-theorem} and \ref{Weak-asymptotic-completeness} that
\beqa
\hil^{\cpf}_{n}(\De)=\hil^+_n(\De),  \label{main-result-one}
\eeqa
where $\hil^+_n(\De)$ is the subspace of $n$-particle Haag-Ruelle scattering states (of particles from $H_m$) with total energy-momentum in $\De$.
Since  the vacuum and the neutral single-particle states are also `charged \pairs free',  we set
\beqa
\hil^{\cpf}:=\complex \Om\oplus \one_{H_m}(U)\hil\oplus \Span\{ \,\hil^\cpf_{n}(\De) \,:\, n\geq 2, \De\subset G_{2m}\,  \}^{\cl}, 
\label{charged-particles-free}
\eeqa
where the span extends over all open bounded sets $\De$ s.t. $(\ov{\De}-\ov{\De})\cap \Sp \, U=\{0\}$ \footnote{If the multiparticle spectrum $G_{2m}$ contains an embedded mass hyperboloid $H_{m'}$, $m'\geq 2m$, the corresponding spectral subspace belongs to the orthogonal complement of  $\hil^{\cpf}$ by relation~(\ref{main-result-one}). This is conceptually not completely satisfactory, since the particles from $H_{m'}$ are  neutral. One could
improve on this point by including also detectors sensitive to particles from $H_{m'}$ and using a variant of Haag-Ruelle theory suitable for embedded
mass-shells \cite{He71,Dy05}. However, we leave this problem for future investigations.     }. Making use of (\ref{main-result-one}), we immediately obtain the generalized AC relation $ \hil^{\cpf}=\hil^+$ and  criterion  (\ref{criterion-for-AC}) for conventional AC.

A crucial technical step of our analysis is the proof of existence of the limits~(\ref{n-detector-limit}). 
We recall that the convergence of Araki-Haag detectors on the subspace of scattering states of bounded energy
follows from the results in \cite{AH67, Bu90}. However, their convergence on the orthogonal complement of this subspace,
which is of great importance for the question of AC, is a long-standing open problem, discussed for example in \cite{Ha}.
To tackle this problem,  we essentially reduce it  to  scattering theory  of 
an $n$-body {\em dispersive Hamiltonian}. Let us explain this reduction:

Let us set $Q_{n,t}(\De):=C_{1,t}\ldots C_{n,t}\one_{\De}(U) $. Exploiting locality and the disjointness
of supports of $h_i$ (in the first variable) we  can write:
\begin{align}
& Q_{n,t}(\De)\Psi=\int \, \H_{t}^{\w}(\hx,\hy)   B_1^*(t,x_1)\ldots  B_n^*(t,x_n)B_1(t,y_1)\ldots  B_n(t,y_n)\Psi d\hx d\hy +O(t^{-\infty}), 
\label{Q-approximants}
\end{align}
where $ \hx:=(x_1,\ldots, x_n)$, $\hy:=(y_1,\ldots, y_n)$ and we denote by $\H_{t}^{\w}(\hx,\hy)$
the distributional kernel of 
\beq\label{tozu}
\H_{t}^{\w}:= h_{1,t}^{\w}\otimes\cdots\otimes h_{n,t}^{\w}
\eeq 
and by $O(t^{-\infty})$ a term which vanishes in norm faster than any
inverse power of $t$.

 Exploiting the fact that $\Psi\in \one_{\De}(U)\hil$ and our assumptions on the energy-momentum transfers of
$B_i$, we can write
\beqa
B_1(t,y_1)\ldots  B_n(t,y_n)\Psi=\Om(\Om| B_1(t,y_1)\ldots  B_n(t,y_n)\Psi)_{\cH}.
\eeqa
We set 
\[
F_t(\hy):=(\Om| B_1(t,y_1)\ldots  B_n(t,y_n)\Psi)_{\cH},
\]
and  note that by a result from \cite{Bu90}, $F_t\in L^2(\real^{nd})$ for any
$t\in \real$.  Thus we obtain from (\ref{Q-approximants}): 
\begin{align}
 Q_{n,t}(\De)\Psi=\int\bigg(\int  \H_{t}^{\w}(\hx,\hy) F_t(\hy)d\hy\bigg)  B_1^*(t,x_1)\ldots  B_n^*(t,x_n)\Om d\hx+O(t^{-\infty}).
\label{almost-Haag-Ruelle}
\end{align}
If we  replaced the expression in bracket above by a sum of products of $n$ positive energy solutions of the Klein-Gordon
equation, the first term on the r.h.s. of   (\ref{almost-Haag-Ruelle}) would become an $n$-particle scattering state approximant.
While such a substitution is not possible at finite times, it can be performed asymptotically: 
In fact, as we show in \Theorem~\ref{Klein-Gordon-approximation},  there exists the limit
\beqa
F_+=\lim_{t\to\infty} \e^{\i t\tom(D_{\hx}) }\H_{t}^{\w}F_t, \label{intermediate-lim}
\eeqa
where $\tom(D_{\hx}):=\om(D_{x_1})+\cdots+\om(D_{x_n})$. In \Theorem~\ref{Main-result-half} we verify that the existence
of this limit implies the convergence of $t\mapsto Q_{n,t}(\De)\Psi$ as $t\to\infty$.
The key step towards the proof of convergence in (\ref{intermediate-lim}), which we take in Lemma~\ref{almost-KG-equation},
is to show that $F_t$ satisfies the following  evolution equation with a source term:
\beqa
\p_{t}F_{t}= -\i \tilde{\omega}(D_{\hx})F_{t}+R_{t}, \label{approx-evol-eq}
\eeqa
where the source term satisfies $\H_{t}^{\w}R_{t}=O(t^{-\infty})$ due to locality and the disjointness of supports of $h_i$ in the first argument. 

It is easy to see that the Schr\"odinger equation of a system of massive particles with relativistic dispersion relations, interacting
with a rapidly decaying potential, has a general form of (\ref{approx-evol-eq}). Thus we reduced the problem of convergence of
the generalized Araki-Haag detectors in (\ref{n-detector-limit}) to the question of existence of the limit~(\ref{intermediate-lim}) in a dispersive system
described by the evolution equation~(\ref{approx-evol-eq}). 

For $n=2$ we solved this problem in a recent publication \cite{DG12}, for
standard Araki-Haag detectors whose symbols $h_i$ are independent of momentum,
following the approach of Graf \cite{Gr90}: 
we combined a large velocity propagation estimate, which in our context says  that particles cannot 
move faster than light, with a phase space propagation estimate, which encodes the fact  that the instantaneous velocity of a particle 
 equals its average velocity at large times. The convex Graf function, appearing in the derivation of this latter estimate, must vanish
near the collision plane $\{\, x_1=x_2\,\}$ to ensure a rapid decay of the rest term  $R_{t}$ in (\ref{approx-evol-eq}). Due to this restriction, 
the method does not generalize to the case $n>2$, which involves several collision planes ($\{\, x_1=x_2\,\},  \{\, x_1=x_3\,\}, \{\, x_2=x_3\,\}$, etc.)
In fact,  since  the Graf function is convex, it would  have to vanish in the convex hull of these collision planes, which contains the relevant part of
the configuration space.  This difficulty is one of several obstacles which hinder our understanding of AC for dispersive systems of three or more 
particles \cite{Zi97,Ge91}.

A solution of this problem in the case of a product of $n\geq 3$ particle detectors is the main technical result of the present paper.
In this case it is instrumental to use symbols $h_i$ in (\ref{one-particle-detector}) which depend also on momentum. As we mentioned above,
they have  the form $h_{i}(x,\xi)=h_{0,i}(x)\chi(x-\nabla\om(\xi))$, where the supports of $h_{0,i}\in C_0^{\infty}(\real^d)$ are
  disjoint 
(with some minimal distance $\eps>0$) and $\chi$ is supported in a ball around  zero whose radius is $\eps'\ll\eps$. For such  symbols $h_{i}$ and $H_{t}^{\w}$ as in (\ref{tozu}) we prove in \Proposition~\ref{novel-propagation} the following new variant of a phase space propagation estimate:
\beq\label{aza}
\int_{1}^{+\infty} \| ( \hx/t- \nabla \ti\omega(D_{\hx}))\cdot H_{ t}^{\w} F_{t}\|^{2}\frac{dt}{t}<\infty.
\eeq
 Abstract arguments, which are an extension of results of standard scattering theory  to inhomogeneous evolution equations like (\ref{approx-evol-eq}), allow then to deduce from (\ref{aza}) the existence of the limit (\ref{intermediate-lim}).

Our paper is organized as follows: In Section~\ref{Framework} we recall the framework of algebraic QFT, introduce some central concepts
and  state our main results.  Section~\ref{sec1} contains  more technical preliminaries. In Section~\ref{echi} we show that the existence
of the intermediate limit~(\ref{intermediate-lim}) implies the convergence of the approximating sequences of detectors in (\ref{n-detector-limit}).
Section~\ref{echa} contains the proof of existence of the intermediate 
limit~(\ref{intermediate-lim}). In Section~\ref{Haag-Ruelle-appendix} we
show that the ranges of the asymptotic observables~(\ref{n-detector-limit})
span the entire subspace of the Haag-Ruelle scattering states.

This paper can be seen as a (non-trivial) generalization of our work \cite{DG12}
on two-particle scattering to the $n$-particle case.  Readers who are familiar
with \cite{DG12} will find material which is special
to the $n>2$ case    in Subsections~\ref{pseudo-subsection}--\ref{first-pseudo-section},
\ref{sec1.4},\ref{almost-prelim} and in Section~\ref{echa}.

\vspace{0.5cm}

\noindent{\bf Acknowledgment:} W.D. would like to thank 
D. Buchholz, W. De Roeck, J. Fr\"ohlich,  G.~M.~Graf, A. Jaffe and J. S. M\o ller  for interesting discussions.
Financial support  of the  German Research Foundation (DFG) within the stipend DY107/1--1 
is gratefully acknowledged.

\section{Framework and results}\init\label{Framework}

In this section we  recall the Haag-Kastler framework of local quantum field theory and state our main results. The preliminary Subsections~\ref{general-assumptions} and \ref{relevant-classes} are similar to the corresponding  subsections of \cite{DG12}.

\subsection{Nets of local observables}\label{general-assumptions}

We base our theory  on  a  net  
\[
\mco\mapsto \mfa(\mco)\subset B(\hil)
\] 
of von Neumann algebras attached to open bounded regions of Minkowski spacetime $\real^{1+d}$,
which satisfies the assumptions of isotony, locality, covariance w.r.t. translations, positivity of energy,
uniqueness of the vacuum and cyclicity of the vacuum. 

\emph{Isotony} says  that $ \mfa(\mco_1)\subset \mfa(\mco_2)$ if $\mco_1\subset\mco_2$, which
 allows to define the $C^*$-inductive limit of the net, denoted by  $\mfa$.
\emph{Locality} requires that $\mfa(\mco_1)\subset \mfa(\mco_2)'$ if $\mco_1$ and $\mco_2$ are spacelike
separated. To state the remaining postulates, we  introduce
a strongly continuous unitary representation of translations 
\[
\real^{1+d}\ni (t,x)\mapsto U(t, x)=: \e^{\i (tH- x\cdot P)}\hbox{ on }\cH,
\]
which induces a group of automorphisms of $\mfa$: 
\[
\alpha_{t,x}(B):= B(t,x):=U(t,x)BU(t,x)^{*}, \ B\in \mfa, \ (t,x)\in \rr^{1+d}.
\]
\emph{Covariance} requires that
\beqa
\alpha_{t,x}(\mfa(\mco))=\mfa(\mco+(t,x)), \ \forall \  \hbox{open bounded } \mco \hbox{ and }  (t,x)\in \rr^{1+d}.
\eeqa
We will need a restrictive formulation of positivity of energy, suitable for massive theories. 
We denote  by $H_m:=\{ (E,p)\in \real^{1+d}\setbar E=\sqrt{p^2+m^2}\}$ the mass hyperboloid of a particle of mass $m>0$
and set $G_{\mu}:=\{ (E, p)\in \real^{1+d}\setbar E\geq \sqrt{p^2+\mu^2}\}$.  We assume that:
\beq\bes\label{spec-ass}
i) \ \ & \Sp\, U=\{0\}\cup H_m\cup G_{2m}, \\ 
ii) \ \ & \one_{\{0\}}(U)=|\Omega\rangle\langle \Omega|, \ \Omega\hbox{ cyclic for }\mfa. 
\ees\eeq
Here we denoted by $\Sp\, U\subset \rr^{1+d}$ 
the spectrum of $(H,P)$ and by $\one_{\Delta}(U)$ the spectral projection on a Borel set $\Delta\subset \rr^{1+d}$. 
The unit vector $\Omega$ will be called the {\em vacuum vector}. 
Part $i)$ in (\ref{spec-ass})  encodes \emph{positivity of energy} and the presence of an upper and lower mass gap $m$.  
Part $ii)$ covers the \emph{uniqueness and cyclicity of the vacuum}.

\begin{remark}
We adopt the restrictive form of the spectrum condition (\ref{spec-ass}) $i)$ to remain consistent with the discussion of AC in the Introduction.
We remark, however,  that our main results, \Theorems~\ref{Main-theorem} and \ref{Weak-asymptotic-completeness} below, remain valid as they stand if the 
assumption~(\ref{spec-ass}) $i)$ is relaxed to $\Sp\, U=\{0\}\cup H_m\cup \ti G$, $\ti G\subset G_{\mu}$, $m<\mu\leq 2m$.  
If $\ti G\backslash G_{2m}$ consists of isolated mass hyperboloids, our results can easily be modified so as to take the additional types
of neutral particles into account. 
\end{remark}

\subsection{Relevant classes of observables} \label{relevant-classes}

In this subsection we introduce some classes of observables, which are important for our discussion. 
We start with the definition of \emph{almost local} operators.  We denote by  $\mco_{r}:=\{\,(t,x)\in \rr^{1+d}\ : \ |t|+|x|<r\,\}$ 
 the double cone of radius $r$ centered at $0$.
\begin{definition}
$B\in \mfa$  is  {\em almost local}  if  there exists a family $A_{r}\in \mfa(\mco_{r})$  s.t.  $\| B- A_{r}\| \in O(\langle r\rangle^{-\infty})$.
\end{definition}

For $B\in \mfa$, we denote by  $\widehat{B}$ the Fourier transform of $(t,x)\mapsto B(t,x)$ defined as an operator-valued distribution:
\begin{equation}
\label{toto.e2}
\widehat{B}(E, p):= (2\pi)^{-(1+d)/2}\int\e^{-\i(Et-p\cdot x)}B(t,x)dtdx.
\end{equation}
The support of $\widehat{B}$,   denoted by $\supp(\widehat B)\subset \rr^{1+d}$, is called the {\em energy-momentum transfer of } $B$. 
We recall  the following well-known properties:
\beq\bes
\label{toto.e3}
i) \  \ & \widehat{\alpha_{t,x}(B)}(E, p)= \e^{\i (Et-p\cdot x)}\widehat{B}(E, p),\\
ii) \ \ & \supp(\wh{B^{*}})= - \supp(\widehat B), \\ 
iii) \ \ & B\one_{\Delta}(U)= \one_{\ov{\Delta+\supp(\widehat B)}}(U)B\one_{\Delta}(U), \ \forall \textrm{ Borel sets } \Delta\subset \rr^{1+d}.
\ees\eeq
For iii) we refer to \cite[Theorem 5.3]{Ar82}.
Now we are ready to define another important class of observables, which are the energy decreasing operators:
\begin{definition}
 $B\in \mfa$ is  {\em energy decreasing} if $\supp(\widehat B)\cap V_{+}=\emptyset$, where $V_{+}:= \{(E, p)\ : \ E\geq |p|\}$
is the closed forward light cone.
\end{definition}
\noindent In the rest of the paper we will work with the following set of observables:
\begin{definition}\label{def-de-L0}
 We denote by $\mcL_{0}\subset \mfa$ the subspace spanned by  the elements $B\in \mfa$ such that:
 \beq\bes\nonumber
i) \ \  &B\hbox{ is energy decreasing}, \ \supp(\widehat B)\hbox{ is compact},\\ 
ii)\ \ &\rr^{1+d}\ni (t,x)\mapsto B(t,x)\in \mfa\hbox{ is }C^{\infty}\hbox{ in norm},\\
iii) \ \ & \p^{\alpha}_{t,x}B(t,x) \hbox{ is almost local  for all }\alpha\in \nn^{1+d}. 
\ees\eeq
\end{definition}
Clearly, if {\it i)} and {\it ii)} hold, then $\p_{t,x}^{\alpha}B(t,x)$ is energy decreasing for any $\alpha\in \nn^{1+d}$. 
 It is easy to give examples of elements of $\mcL_0$: let $A\in \mfa(\mco)$ and $f\in \cS(\rr^{1+d})$ with $\supp \widehat{f}$ compact and $\supp\widehat{f}\cap V_{+}=\emptyset$.  Then
\beq\label{arita}
B=(2\pi)^{-(1+d)/2} \int f(t,x)A(t,x)dtdx
\eeq
belongs to $\mcL_{0}$, since $\widehat{B}(E,p)= \widehat{f}(E, p)\widehat{A}(E,p)$. (See (\ref{toto.e1}) below for definition of $\widehat f$).

\subsection{Pseudo-differential operators} \label{pseudo-subsection}

We consider the phase space $T^{*}\real^{\n}= \real^{\n}\times (\real^{\n})'$, whose elements are denoted by $(x, \xi)$.  
For $h\in \cS(T^{*}\real^{\n})$ we define  its {\em Weyl quantization} $h^{\w}$ by
\beq
h^{\w} u(x)  = (2\pi)^{-\n}\int \e^{\i (x-y)\cdot \xi}h\big(\frac{x+y}{2}, \xi\big)u(y)  dy d\xi, \ u\in \cS(\real^{\n}).
\label{Weyl-quantization}
\eeq 
It is well known that $h^{\w}$ is bounded on $\cS(\real^{\n})$ and $L^{2}(\real^{\n})$.

Denoting by $A(x, y)\in \cS'(\real^{\n}\times \real^{\n})$ the distributional kernel of $A: \cS(\real^{\n})\to \cS'(\real^{\n})$, one has:
\beq\label{titi}
h^{\w}(x, y)= (2\pi)^{-\n/2}\widecheck{h}\big(\frac{x+y}{2}, x-y\big),
\eeq
where $ \widecheck{h}(x, y)= (2\pi)^{-\n/2}\int \e^{\i y\cdot \xi}h(x, \xi)d\xi $ is the inverse Fourier transform  of $h$ in the $\xi$ variable.

We refer to \cite{Ho} and \cite[Appendix D]{DG97} for
systematic expositions of the Weyl quantization. Properties needed in the present work are summarized in 
Subsection~\ref{first-pseudo-section} below.

\subsection{Results} \label{results-subsection}
To   $B\in \mcL_{0}$, $h\in  \cS(T^{*}\real^{d})$, we  associate the {\em one-particle detector}:
\beq\bes
\label{toto}
C_{t}&:=\int B^{*}(t, x)h_{t}^{\w}(x, y)B(t, y)dxdy\\ 
&=\int B^{*}\left(t, x +\frac{y}{2}\right)\widecheck{h}\left(\xt, y\right)B\left(t, x-\frac{y}{2}\right)dxdy,
\ees\eeq
where we set $h_{t}(x, \xi)= h(\xt, \xi)$. In view of Lemma~\ref{existence-lemma} below, one has
\[
\sup_{t\in\real}\|C_{t} \one_{\ti\Delta}(U)\|<\infty,
\]
 for any bounded Borel set $\ti\De$.

A  much more convenient formula for $C_t$, using notation introduced below in Sect.  \ref{sec1}, is:
\beq
C_{t}= \e^{\i t H}\left(a_{B}^{*}\circ (\one_{\cH}\otimes h_{t}^{\w})\circ a_{B}\right)\e^{-\i tH}. \label{generalized-detector-formula}
\eeq
In particular, it  remains meaningful if  $h_{t}^{\w}$ is replaced by any bounded operator on $L^2(\real^d)$. 
For example, for symbols $h_0\in C_0^{\infty}(\real^d)$, independent of momentum, we recover from (\ref{generalized-detector-formula})
the conventional Araki-Haag detectors from \cite{AH67}:
\[
C_{t}^{\mathrm{AH}}:= \int B^{*}(t,x)h_0\left(\xt\right)B(t,x)dx, 
\]
which we considered  in \cite{DG12}. These detectors are
only sensitive to average velocity $x/t$ of a particle. 
In view of formula~(\ref{toto}),  our detectors $C_{t}$ are essentially averages (w.r.t. $y$) of the conventional Araki-Haag detectors, 
and are also sensitive  to momentum  $\xi$  of a particle.

 We fix now $B_{i}\in \mcL_{0}$, $h_{i}\in \cS(T^{*}\rr^{d})$ for $1\leq i\leq n$ and define $C_{i,t}$ as in (\ref{toto}).  For 
 any  open bounded subset $\De\subset G_{2m}$ we define the {\em n-particle detector}:
\beqa
\C_{n,t}(\De):=C_{1,t}\ldots C_{n,t}\one_{\Delta}(U).
\eeqa
Our main technical result is the strong convergence of $\C_{n,t}(\De)$ as $t\to\infty$ if the extension of $\De$ is smaller than the
mass gap (i.e., $(\ov\De-\ov\De)\cap\Sp\, U=\{0\}$), $B$ is $\Delta-$admissible in the sense of 
\Definition~\ref{delta-admissible} and $\H:= h_1\otimes \cdots \otimes h_n$ is admissible in the sense of \Definition~\ref{support-admissible}.
%%%%%%%%%%%%%%%%%%%%%%%%%%%%%%%%%
\begin{definition}\label{delta-admissible}
 Let $\Delta\subset \rr^{1+d}$ be an open bounded set 
 and $B_{1},\ldots, B_{n}\in \mcL_{0}$. We say that $\un B=(B_{1},\ldots, B_{n})$ is $\Delta-${\em admissible} if 
 \beq\label{transfer-to-hyperboloid}
& (-\supp(\widehat B_{i})) \cap \Sp\, U\subset H_m,\ i=1,\ldots, n,\\[2mm]
& -(\supp(\widehat B_{1})+\cdots+\supp(\widehat B_{n}) )\subset \De, \label{transfer-from-vacuum}\\[2mm]
& (\ov{\De}+\supp(\widehat B_{1})+\cdots+\supp(\widehat B_{n}))\cap \Sp\, U\subset \{0\}. \label{transfer-to-vacuum}
\eeq
\end{definition}
%%%%%%%%%%%%%%%%%%%%%%%%%%%%%%%%%%%%%%%%%
\begin{remark}
In  Lemma~\ref{O-lemma} it is shown that if  $\De\subset G_{2m}$ is an open bounded set s.t. $(\ov\Delta-\ov \Delta)\cap \Sp\, U\subset \{0\}$ and  $-\supp(\widehat B_{1}),\ldots,-\supp(\widehat B_{2})$ are sufficiently small neighbourhoods 
of vectors $\tilde p_1, \ldots, \tilde p_n\in H_m$ s.t. $ \tilde p_i\neq  \tilde p_j$ for $i\neq j$ and $\tilde p_1+\cdots+\tilde p_n\in \De$ then 
$\un B=(B_{1},\ldots, B_{n})$ is $\Delta-$admissible. We also note that for such $\De$ (\ref{transfer-to-hyperboloid}) and (\ref{transfer-from-vacuum})
cannot be simultaneously satisfied for $n=1$ if $B_1\neq 0$.
\end{remark}
Let us introduce the notation
\begin{align}
& \DD_{0}:=\{\, \hx\in \real^{nd}\,:\, x_i=x_j\, \textrm{ for some } i\neq j\,\}, \label{diagonal} \\
&\ball(0, \epsilon):= \{\,\hx\in \rr^{nd}\, : \, |\hx|<\epsilon\}, \\
&\tiball(0, \epsilon):=\{\,\hx\in \rr^{nd}\, : \, |x_i|<\epsilon,\, i=1,\ldots,n\,\},
\end{align}
and define $\ti\om(\un\xi):=\om(\xi_1)+\cdots+\om(\xi_n)$. Note that $\nabla\ti\omega(\un \xi)\in \tiball(0, 1)$ for any $\un \xi\in \real^{nd}$.
%%%%%%%%%%%%%%%%%%%%%%%%%%%%%%%%%%%%%%%%%%%%%%%%%%%%%%%%%%%%%%%%%%%%%%%%%%

\begin{definition}\label{support-admissible} Let $\H\in C_0^{\infty}(T^{*}\real^{nd}; \real^\ell)$. We say that $\H$ is {\em admissible}, if  there exists $K\Subset \real^{nd} $  and $\epsilon>0$ such that $K+ \tiball(0, \epsilon)\subset  \tiball(0,1)   \backslash \DD_{0}$ and
\begin{align}
\supp\, \H \subset \{\, (\hx,\hxi)\in T^{*}\real^{nd}  \, : \, \nabla\ti\omega(\un \xi)\in K, \  \un x- \nabla\ti \omega(\un \xi) \in \ball(0,\epsilon)\, \}.
\label{}
\end{align}
\end{definition}

%%%%%%%%%%%%%%%%%%%%%%%%%%%%%%%%%%%%%%%%%%%%%%%%%%%%%%%%%%%%%%%%%%%%%%%%%%%%%%%
\bet\label{Main-theorem}  Let $\De\subset G_{2m}$ be an open bounded set s.t. $(\ov\De-\ov \De)\cap\Sp\, U=\{0\}$. 
Let $\un B=(B_1,\ldots, B_n)$ be a collection of elements of $\mcL_0$ s.t. $\un B$ is $\De-$admissible and let 
$\un h=(h_1,\ldots, h_n)$ be a collection of elements of $C_0^{\infty}(T^{*}\real^{d})$ s.t. $\H=h_1\otimes\cdots \otimes h_n$ is admissible. 
Then there exists the limit
\beqa
\C_n^+(\De):=\slim_{t\to\infty}C_{1,t}\ldots C_{n,t} \one_{\Delta}(U), \label{main-approximants}
\eeqa
where $C_{i,t}$ are defined in  (\ref{toto}) for  $B_i,h_i$, $i=1,\ldots, n$. The range of $\C_n^+(\De)$
belongs to $\hil_n^+(\De):=\one_{\Delta}(U)\hil_n^+$, where $\hil_n^+$ is the subspace of $n$-particle scattering states.
(See \Definition~\ref{main-Haag-Ruelle-concepts}). 
\eet
%%%%%%%%%%%%%%%%%%%%%%%%%%%%%%%%%%%%%%%%%%%%%%%%%%%%%%%%%%%%%%%%%%%%%%%%%%%%%%%%%%
\proof  Follows immediately from \Theorems~\ref{Main-result-half} and \ref{Klein-Gordon-approximation}. \qed\\
%%%%%%%%%%%%%%%%%%%%%%%%%%%%%%%%%%%%%%%%%%%%%%%%%%%%%%%%%%%%%%%%%%%%%%%%%%%%%%%%%%
\Theorem~\ref{Main-theorem} substantiates our discussion below  formula~(\ref{n-detector-limit}) in the Introduction, where we argued
that vectors from the ranges of $\C_n^+(\De)$ should describe configurations of $n$ neutral particles with total energies-momenta in $\De$. 
This theorem allows us to define the $n$-particle component of the `charged particles free' subspace $\hil^\cpf_{n}(\De)$ associated with any open
bounded  set $\De\subset G_{2m}$
s.t. $(\ov\De-\ov \De)\cap\Sp\, U=\{0\}$:
\beqa
\hil^\cpf_{n}(\De):=\Span\{ Q^+_{n,\al}(\De)\hil(\De)\,:\, \al\in J\,\}^{\cl}, \label{charged-particles-free-one}
\eeqa
where   $J$ is the collection of pairs $\al=(\un B,\un h)$ satisfying the conditions from Thm.~\ref{Main-theorem} and
$\C^+_{n,\al}(\De)$ is the limit (\ref{main-approximants}) corresponding to $\al$. We also define the total
`charged particles free' subspace: 
\beqa
\hil^{\cpf}:=\complex \Om\oplus \one_{H_m}(U)\hil\oplus \Span\{ \,\hil^\cpf_{n}(\De) \,:\, n\geq 2, \De\subset G_{2m}\,  \}^{\cl}, 
\label{charged-particles-free-total}
\eeqa
where the span extends over all open bounded sets $\De\subset G_{2m}$ s.t. $(\ov{\De}-\ov{\De})\cap \Sp \, U=\{0\}$.
It follows immediately from \Theorem~\ref{Main-theorem} that 
\beqa
\hil^\cpf_{n}(\De)\subset \hil_n^+(\De) \textrm{ and
therefore } \hil^{\cpf}\subset \hil^+, \label{AC-inclusions}
\eeqa
where $\hil^+$ is the subspace of all scattering states of neutral particles of mass $m$ (see \Definition~\ref{main-Haag-Ruelle-concepts}). 
The last inclusion can be interpreted as a weak variant of AC, as it says that certain subspace $\hil^{\cpf}\subset \hil$,
defined without reference to scattering states, is in fact contained in $\hil^+$.  The larger the subspace $\hil^{\cpf}$ is, the closer we
are to verifying AC proper. For example, if we could show that $\hil^{\cpf}=\hil$, conventional AC would follow, which gives 
one implication in our criterion for AC stated in (\ref{criterion-for-AC}).  The opposite implication is given by the following theorem,
which shows that  the inclusions in (\ref{AC-inclusions}) are in fact equalities. This result, whose proof is given in  Sect. \ref{blit}, guarantees, in
particular, that $\hil^\cpf_{n}(\De)\neq 0$ for any  $\De$ as specified above.
%%%%%%%%%%%%%%%%%%%%%%%%%%%%%%%%%%%%%%%%%%%%%%%%%%%%%%%%%%%%
\bet\label{Weak-asymptotic-completeness}  
Let $\De\subset G_{2m}$ be an open bounded set such that  $(\ov{\De}-\ov{\De})\cap\Sp\,U= \{0\}$. 
Let $\hil^\cpf_{n}(\De)$ and $\hil^{\cpf}$ be as defined in (\ref{charged-particles-free-one}) and (\ref{charged-particles-free-total}), respectively.
Then
\beqa
\hil^\cpf_{n}(\De)=\hil_n^+(\De) \textrm{ and
therefore } \hil^{\cpf}=\hil^+, \label{AC-inclusions-one}
\eeqa
where $\hil_n^+(\De):=\one_{\Delta}(U)\hil_n^+$, $\hil_n^+$ is the 
subspace of $n$-particle scattering states and $\hil^+$ is the subspace of
all scattering states. (See \Definition~\ref{main-Haag-Ruelle-concepts}). 
\eet
%%%%%%%%%%%%%%%%%%%%%%%%%%%%%%%%%%%%%%%%%%%%%%%%%%%%%%%%%
%%%%%%%%%%%%%%%%%%%%%%%%%%%%%%%%%%%%%%%%%%%%%%%%%%%%%%%%%%%%

%%%%%%%%%%%%%%%%%%%%%%%%%%%%%%%%%%%%%%%%%%%%%%
\section{Preliminaries}\init\label{sec1}
%%%%%%%%%%%%%%%%%%%%%%%%%%%%%%%%%%%%%%%%%
In this section we specify our notation and collect some basic properties of particle detectors. Subsections~\ref{sec1.2}, \ref{particle-detectors}
are similar to \cite[Subsect. 3.2, 3.3]{DG12}. The remaining subsections contain essential generalizations of the material from \cite{DG12}.

\subsection{Notation}\label{sec1.1}
\begin{enumerate}
\item[-] By $x,x_1,x_2,\ldots $ we  denote elements of $\real^d$ and by $\xi, \xi_1,\xi_2, \ldots $ elements of $(\real^d)'$.
We write $T^*\real^d:=\real^d\times (\real^d)'$ to denote the phase space.
  
\item[-] We  set $\hx=(x_1,\ldots,x_n)$ and $\hxi=(\xi_1,\ldots,\xi_n)$ to denote elements of $\real^{nd}$ and $(\real^{nd})'$.
The Lebesgue measure on $\real^{nd}$ is denoted $d\hx$.

\item[-] We write $K\Subset \rr^{\ell}$ if $K$ is a compact subset of $\rr^{\ell}$.

\item[-] We set $\langle x\rangle:= (1+ x^{2})^{\12}$ for $x\in \rr^{d}$ and $\omega(p)= (p^{2}+ m^{2})^{\12}$ for $p\in \rr^{d}$.

\item[-] We denote the momentum operator $\i^{-1}\nabla_{x}$ by $D_{x}$.

\item[-] We denote by $(t, \rx)$ or $(E, p)$  elements of $\rr^{1+d}$.

\item[-] If $f: \rr^{1+d}\to \cc$ we will denote by $f_{t}: \rr^{d}\to \cc$ the function $f_{t}(\,\cdot\,):= f(t,\,\cdot\,)$.

\item[-] We denote by $\cS(\rr^{1+d})$ the Schwartz class in $\rr^{1+d}$. If $f\in \cS(\rr^{1+d})$ we define its (unitary) Fourier transform:
\beq\bes
\label{toto.e1}
\widehat{f}( E, p)&:=(2\pi)^{-(1+d)/2}\int \e^{\i(Et-p\cdot x)}f(t,x)dt dx, \\[2mm]
\widecheck{f}( t, x)&:=(2\pi)^{-(1+d)/2}\int \e^{-\i(Et-p\cdot x)}f(E,p)dE dp.
\ees\eeq
(Note the different sign in the exponent in comparison with (\ref{toto.e2}), where the Fourier transform is taken in the sense of operator valued distributions).

\item[-] If $f\in \cS(\rr^{d})$ we set, consistently with (\ref{toto.e1}),
\begin{align}
\begin{split}
\label{toto.e1-e1-e1}
\widehat{f}(p)&:= (2\pi)^{-d/2}\int\e^{-\i p\cdot x}f(x)dx, \\
\widecheck{f}(x)&:= (2\pi)^{-d/2}\int \e^{\i p\cdot x }f(p)dp.
\end{split}
\end{align}

\item[-] If $h\in \cS(T^*\rr^{d})$ is a symbol, $\widehat{h}$ and $\widecheck{h}$ denote the Fourier transform
and the inverse Fourier transform \emph{w.r.t. the momentum variable $\xi$ only}.  
\item[-] By $\pi_x: T^*\rr^{d}\to \rr^d$
we will denote the projection from the phase space to configuration space.

\item[-] If $B$ is an observable, we write $B^{(*)}$ to denote either $B$ or $B^{*}$.  We will also set 
\[
B_{t}:= B(t, 0), \ B(x):= B(0,x)\hbox{ so that }B(t,x)= B_{t}(x).
\]

\end{enumerate}

\subsection{Pseudodifferential calculus} \label{first-pseudo-section}

For future reference, we recall the following well-known facts:
%%%%%%%%%%%%%%%%%%%%%%%%%%%%%%%%%%%%%%%%%%%%%%%%%
\bep\label{pseudodifferential} Let $h\in \cS(T^{*}\rr^{\n})$, $h_t(x,\xi):=h(x/t,\xi)$ and $\om\in C^{\infty}(\real^\n)$  such that $\p^{\alpha}_{\xi}\omega$ is bounded for all $|\alpha|\geq 1$. Then
\begin{enumerate}[label=\textup{(\arabic*)},ref=\textup{\arabic*},leftmargin = *] 

\item\label{positivity} If $h\geq 0$ then $h_t^\w\geq -\fr{c}{t}$ for some $c\geq 0$ independent of $t$.

\item\label{limit} $\slim_{t\to \infty}\e^{\i t\om(D_x)} h_t^\w \e^{-\i t\om(D_x)}=h(\nabla\om(D_x),D_x)= h(\nabla \omega(\cdot), \cdot)^{\w}$.

\item\label{commutator} $[\om(D_x), \i h_t^\w]$, defined as a continuous linear map $\cS(\real^\n)\to \cS(\real^\n)$,  extends to 
a bounded operator on $L^2(\real^\n)$ s.t.: 
\beqa
[\om(D_x),\i h_t^\w]=\fr{1}{t}\big(\nabla_{\xi}\om\cdot (\nabla_xh)_t \big)^\w+O(t^{-2}).  \non
\eeqa

\item\label{polynomials} 
Let $f\in C^{\infty}(\real^\n)$ be bounded by a fixed polynomial, together with all its derivatives.
Then $f(x/t) h^\w_t$ and $f(D_x) h^\w_t$, defined as continuous linear maps $\cS(\real^\n)\to \cS(\real^\n)$, extend
to bounded operators on $L^2(\real^\n)$ s.t.
\beqa
f(x/t) h^\w_t=(h_{1,t})^{\w}+O(t^{-1}), \quad f(D_x) h^\w_t=h_{2,t}^\w+O(t^{-1}),\non
\eeqa  
where $h_{1,t}(x,\xi):=f(x/t)h(x/t,\xi)$ and $h_{2,t}(x,\xi)=f(\xi)h(x/t,\xi)$.

\item\label{product} Let $g\in \cS(T^*\real^\n)$ and $g_t(x,\xi)=g(x/t,\xi)$. Then $(g_t^\w)(h_t^\w)=(g_t h_t)^\w+O(t^{-1})$.

\item\label{supports} Let $\chi\in C_0^{\infty}(\real^\n)$ be s.t. $\chi(x)=1$ near $\pi_{x}\supp\, h$. Let $\chi_t(x)=\chi(x/t)$ and denote the corresponding operator on $L^2(\real^\n)$ also by $\chi_t$. Then $(1-\chi_t)h_t^\w=O(t^{-\infty})$.

\end{enumerate}
\eep

%%%%%%%%%%%%%%%%%%%%%%%%%%%%%%%%%%%%%%%%%%%%%%%%%%

\subsection{Auxiliary maps \texorpdfstring{$\mathbf{a_B}$}{Auxiliary maps} } \label{sec1.2}  

For $B\in\mfa$, $f\in\cS(\rr^{d})$ we write:
\[
B(f):= \int B(x)f(x)dx,
\] 
so that $B^{*}(f)= B(\overline{f})^{*}$.  If $B_{1}, B_{2}\in \mfa$ are almost local, then
\begin{equation}
\label{quasi}
\| [B_{1}(x_{1}), B_{2}(x_{2})]\|\leq C_{N}\langle x_{1}- x_{2}\rangle^{-N}, \ \forall\ N\in \nn, 
\end{equation}
and consequently
\begin{equation}
\label{quasi-1}
\|[B_{1}(f_{1}), B_{2}(f_{2})]\| \leq C_{N}\int |f_{1}(x_{1})|\langle x_{1}- x_{2}\rangle^{-N}|f_{2}(x_{2})| dx_{1}dx_{2}, \ f_{1}, f_{2}\in \cS(\rr^{d}).
\end{equation}
Now we introduce  auxiliary maps which will be often used  in our investigation: 
\bed
 Let $B\in\mfa$. We denote by $a_{B}: \cH \to \cS'(\rr^{d}; \cH)$ the linear operator defined as:
\[
a_{B}\Psi(x):= B(x)\Psi, \ x\in \rr^{d}.
\]
\eed

\noindent The operator $a_{B}: \cH \to  \cS'(\rr^{d}; \cH)$ is continuous and
\beq\label{toto.e01}
 B(f)= (\one_{\cH}\otimes\langle \overline{f}|)\circ a_{B}, \ f\in \cS(\rr^{d}),
\eeq
where $(\one_{\cH}\otimes\langle \overline{f}|): \cS'(\rr^{d}; \cH)\to \hil$ is defined on simple tensors by
\beqa
(\one_{\cH}\otimes\langle \overline{f}|)(\Psi\otimes T)=T(f)\Psi, \ \Psi\in \hil, T\in  \cS'(\rr^{d}).
\eeqa 
By duality $a_{B}^{*}: \cS(\rr^{d}; \cH)\to \cH$ is continuous and we have
\beq\label{toto.e02}
B^{*}(f)= a_{B}^{*}\circ (\one_{\cH}\otimes |f\rangle), \ f\in \cS(\rr^{d}).
\eeq
The group of space translations
\[
\tau_{y}\Psi(x):= \Psi(x-y), \ y\in\rr^{d},
\]
is  strongly continuous  on  $\cS'(\rr^{d}; \cH)$, and its generator is $D_{x}$ i.e.,
$\tau_{y}= \e^{- \i y\cdot D_{x}}$. Clearly, we have the identity:
\begin{equation}
\label{toto.e0}
a_{B}\circ\e^{-\i y\cdot P}= \e^{-\i y\cdot (D_{x}+ P)}\circ a_{B}, \ y\in \rr^{d}.
\end{equation}
The following lemma  collects  some elementary properties of $a_{B}$.
\begin{lemma}\label{newlemma}
 Let $B\in\mfa$. Then:
 \ben \item For any Borel set $\Delta\subset \rr^{1+d}$:
 \beq\bes\non
a_{B}\one_{\Delta}(U)&=(\one_{\ov{\Delta+ \supp(\widehat B)}}(U) \otimes \one_{\cS'(\real^d)})\circ a_{B}\one_{\Delta}(U), \\[2mm]
a_{B}^{*}\circ(\one_{\Delta}(U)\otimes \one_{\cS(\real^d)})&=\one_{\ov{\Delta-\supp(\widehat B) }}(U)a_{B}^{*}\circ(\one_{\Delta}(U)\otimes \one_{\cS(\real^d)}).
\ees\eeq
\item For any  $f\in \cS(\rr^{d})$ one has $f(D_{x})a_{B}= a_{B_{f}}$ for
\beq\bes\non
B_{f}:=(2\pi)^{-d/2}\int \widecheck{f}(-y)B(0, y)dy
 = (2\pi)^{-(1+d)/2}\int f(-p)\widehat{B}(E, p)dE dp.
 \ees\eeq
Moreover, $\widehat{B_{f}}(E, p)=f(-p)\widehat{B}(E, p)$.
\item If $\supp(\widehat B)$ and $\De$ are compact and $f\in C^{\infty}(\rr^{d})$ then  $f(D_{x})a_{B}\one_{\Delta}(U)= a_{B_{f}}\one_{\Delta}(U)$.
\een
\end{lemma}
\proof (1) follows from (\ref{toto.e3}), (2) is an easy consequence of~(\ref{toto.e0}) and (3) follows from (1) and (2). \qed

\medskip

The mappings $a_{B}$ have much stronger properties if $B\in \mcL_{0}$. For example,  for $\Delta\Subset \rr^{1+d}$  
the operator $a_{B}\one_{\Delta}(U)$
maps $\cH$ into $L^{2}(\rr^{d}; \cH)\simeq \cH\otimes L^{2}(\rr^{d})$, as shown in Lemma \ref{toto.1} below. 
This is a consequence of the  following important property of  energy decreasing operators, proven in~\cite{Bu90}.
%%%%%%%%%%%%%%%%%%%%%%%%%%%%%%%%%%%%%%%%%%%%%%%%%%%%%%%%%%%%%%%%%%%%%%%%%%%%%%%%%
\bel\label{existence-lemma} Let $B\in \mfa$ be energy decreasing   with $\supp(\widehat B)\Subset \rr^{1+d}$ 
and $\De\subset\real^{1+d}$ be some bounded Borel set. 
Then there exists $c\geq 0$ such that for any  $F\Subset \real^d$  one has:
\beqa
\|\int_{F}  (B^*B)(x)\one_{\Delta}(U)dx\|\leq c\int_{F- F} \|[B^*,B(x)]\|dx. \label{harmonic}
\eeqa
\eel
%%%%%%%%%%%%%%%%%%%%%%%%%%%%%%%%%%%%%%%%%%%%%%%%%%%%%%%%%%%%%%%%%%%%%%%%%%%%%%%%%%
Note that if $B$ in Lemma~\ref{existence-lemma} is in addition almost local 
then the function $\real^d\ni x\mapsto \|[B^{*}, B(x)]\|$ vanishes faster than any inverse power of $|x|$ as $|x|\to \infty$, 
hence we can take $F=\real^d$ in (\ref{harmonic}).
In view of Lemma~\ref{existence-lemma}, it is convenient to introduce the subspace of vectors with compact energy-momentum spectrum:
\[
\cH_{\rm c}(U):= \{\Psi\in \cH \ : \ \Psi= \one_{\Delta}(U)\Psi, \ \Delta \Subset \rr^{1+d}\}. 
\]
There holds the following simple fact:

\begin{lemma}\label{toto.1} Assume that  $\Delta\Subset \rr^{1+d}$ and let $B\in \mcL_{0}$. Then
\[
a_{B}\one_{\Delta}(U): \cH\to \cH \otimes L^{2}(\rr^{d})\hbox{ is bounded.}
\] 
\end{lemma}

\proof We note that
\[
\one_{\Delta}(U)a_{B}^{*}\circ a_{B}\one_{\Delta}(U)= \int_{\rr^{d}}\one_{\Delta}(U)(B^{*}B)(\rx) \one_{\Delta}(U) d \rx,
\]
and apply  Lemma \ref{existence-lemma}.   \qed

\begin{remark}
Considering  $a_{B}$ as a linear operator from $\cH$ to $ \cH\otimes L^{2}(\rr^{d})$ with domain $\cH_{\rm c}(U)$, we see that $\cH\otimes \cS(\rr^{d})\subset \Dom a_{B}^{*}$, hence $a_{B}$ is closable.
\end{remark}
%%%%%%%%%%%%%%%%%%%%%%%%%%%%%%%%%%%%%%%%%%%%%%%
\subsection{Particle detectors} \label{particle-detectors}
In this subsection we establish connection between the maps $a_B$ and  the particle detectors $C_t$ introduced in (\ref{toto}).

\begin{definition}\label{toto.2}
 Let $B\in \mcL_{0}$. For $h\in B(L^{2}(\rr^{d}))$ we set:
 \[
N_{B}(h):= a_{B}^{*}\circ(\one_{\cH}\otimes h)\circ a_{B}, \ \Dom N_{B}(h)= \cH_{\rm c}(U).
\]
\end{definition}
From  Lemmas \ref{newlemma} and  \ref{toto.1} we obtain the following facts:

\begin{lemma}\label{toto.3} We have:
\beq\bes\non
(1) \ \ &\|N_{B}(h)\one_{\Delta}(U)\|_{B(\cH)}\leq c_{\Delta, B}\| h\|_{B(L^{2}(\rr^{d}))},\\ 
(2) \ \ & \forall \ \Delta\Subset \rr^{1+d}, \ N_{B}(h)\one_{\Delta}(U)= \one_{\Delta_{1}}(U)N_{B}(h)\one_{\Delta}(U), \hbox{ for some }\Delta_{1}\Subset \rr^{1+d}.
\ees\eeq
\end{lemma}

\noindent Denoting by $h(\rx,y)$ the distributional kernel of $h$ we have the  following  expression for $N_{B}(h)$,
\beq\label{toto.e1bis}
N_{B}(h)= \int B^{*}(x)h(x, y) B(y)d\rx dy,
\eeq
which is meaningful as a quadratic form identity on $\cH_{\rm c}(U)$.  This shows that for $h\in \cS(T^{*}\rr^{d})$, 
$h_{t}(x,\xi):= h\left(\xt,\xi \right)$, $B_{t}:= B(t, 0)$ and  $C_{t}$  as
defined in (\ref{toto}), one has:
\[
C_{t}= N_{B_t}(h_{t}^{\w}).
\]

\subsection{Auxiliary maps \texorpdfstring{${\mathbf a_{\un B} }$}{Auxiliary maps} }\label{sec1.4}
 We recall that $\un B=(B_1,\ldots, B_n)$, $B_i\in \mcL_0$ and $\un x=(x_1,\ldots, x_n)$, $x_i\in\real^d$.
The Lebesgue measure in $\real^{nd}$ is denoted $d\un x$.
We state the following definition which is meaningful
due to Lemma~\ref{toto.1}:

\bed For $B_{1},\ldots, B_{n}\in \mcL_{0}$ we  define the linear operator:

\begin{equation}
\label{toto.e5}
a_{\BB}: \ \begin{array}{rl}
&\cH_{\rm c}(U)\to  \cH\otimes L^{2}(\rr^{nd},d\hx  ),\\[2mm]
&\Psi\mapsto a_{\BB}\Psi= (a_{B_{1}}\otimes \one_{L^{2}(\rr^{(n-1)d})} ) \circ\cdots\circ 
 (a_{B_{n-1}}\otimes \one_{ L^{2}(\rr^{d}) } )  \circ a_{B_{n}}\Psi.
\end{array}
\end{equation}
\eed

\noindent Formally we have 
\[
a_{\BB}\Psi(\rx_{1},\ldots, \rx_{n})= B_{1}(\rx_{1})\ldots B_{n}(\rx_{n})\Psi.
\]
We state the following lemma, which is a direct consequence of Lemmas \ref{newlemma} and \ref{toto.1}.

\begin{lemma}\label{toto.-1}
 Let $\Delta\Subset \rr^{1+d}$  and let $B_{1},\ldots, B_{n}\in \mcL_{0}$. Then:
\ben
\item $a_{\BB}\one_{\Delta}(U): \cH\to  \cH\otimes L^{2}(\rr^{nd},d\hx )$ is bounded,

\item  For any $\Delta \Subset \rr^{1+d}$ one has:
\beq\bes\non
&a_{\BB}\one_{\Delta}(U)=( \one_{\Delta+\supp(\widehat B_{1})+\cdots+\supp(\widehat B_{n}) }(U)\otimes \one_{L^{2}(\real^{nd}) })\circ a_{\BB}\one_{\Delta}(U), \\[2mm]
& a_{\BB}^{*}\circ ( \one_{\Delta}(U)\otimes \one_{L^{2}(\real^{nd})} )= 
\one_{\Delta-(\supp(\widehat B_{1})+\cdots+\supp(\widehat B_{n}))  }(U) a^{*}_{\BB}\circ  ( \one_{\Delta}(U)\otimes \one_{L^{2}(\real^{nd}) }).
\ees\eeq
\een
\end{lemma}

\subsection{Some consequences of almost locality} \label{almost-prelim}

We collect some commutator estimates which make essential use of almost locality.  The proofs are  given in Appendix \ref{stup-proof}. 

It is convenient to introduce the following functions for $N >d$:
\beq\label{toto.e13bis}
g_{N}(\xi):= \int \e^{- \i \xi\cdot \rx} \langle \rx\rangle^{-N}d \rx.
\eeq
Clearly 
\[
\p_{\xi}^{\alpha}g_{N}(\xi)\in O(\langle \xi\rangle^{-p}), \ \forall \ p\in\nn, \ |\alpha|< N- |d|,
\]
and the operator on $L^{2}(\rr^{d})$ with kernel $\langle x- y\rangle^{-N}$ equals $g_{N}(D_{x})$.
%%%%%%%%%%%%%%%%%%%%%%%%%%%%%%%%%%%%%%%%%%%%%%%%%%%%%%%%%%%%%%%
\begin{proposition}\label{stup} Let  $h_{i}\in B(L^{2}(\rr^{d}))$,  $B_{i}\in \mcL_{0}$, $1\leq i\leq n$. We set $\un B=(B_1,\ldots, B_n)$,  
$\ov B= (B_{n}, \ldots , B_{1})$  and 
 \beq
R_{n}:= a^{*}_{\ov B}\circ (\one_{\cH}\otimes h_{1}\otimes\cdots \otimes h_{n})\circ a_{\un B}
-\prod_{i=1}^{n}N_{B_{i}}(h_{i}).  \label{R-n-definition}
\eeq 
Let us fix  measurable functions $\chi_{i}:\rr^{d}\to \rr$ with $0\leq \chi_{i}\leq 1$ and still denote by $\chi_{i}\in B(L^{2}(\rr^{d}))$ the operator of multiplication by $\chi_{i}(x)$.

Then  for any  $\Delta\Subset \rr^{1+d}$, $N\in\nn$ there exists a constant $C_{N}(\Delta, B_{1}, \dots, B_{n})$ such that:
 \beq\bes\non
\|R_{n}\one_{\Delta}(U)\|_{B(\cH)}
\leq & \, C_{N}(\Delta, B_{1}, \dots, B_{n})\times \\ 
&\times\bigg(   \sum_{i=1}^{n} \big(\| h_{i}(1- \chi_{i})\|_{B(L^{2}(\real^d) )}+ \| (1- \chi_{i})h_{i}\|_{B(L^{2}(\real^d) )}\big)
\prod_{j\neq i}\| h_{j}\|_{B(L^{2}(\real^d) )} \\
&\quad+\sum_{i\neq j} \| \chi_{i}g_{N}(D_{x})\chi_{j}\|_{B(L^{2}(\real^d))}\prod_{i=1}^{n}\| h_{i}\|_{B(L^{2}(\real^d) ) }   \bigg).
\ees\eeq 
\end{proposition}
%%%%%%%%%%%%%%%%%%%%%%%%%%%%%%%%%%%%%%%%%%%%%%%%%%%%%%%%%%%%%%%%
\begin{remark}
 Let us explain the meaning of \Proposition~\ref{stup}. By almost locality we expect $R_{n}$ to be small if the operators $h_{i}$ are `localized' in distant regions of configuration space. This is easy to prove if   $h_{i}= h_{i}(x)$ for functions $h_{i}$ with compact, pairwise disjoint supports.
In the general case we pick functions $\chi_{i}$ such that the operators $h_{i}(1- \chi_{i})$ and $(1-\chi_{i})h_{i}$ are small, i.e., $h_{i}$ is `localized' in the support of $\chi_{i}$. If these supports are far away from each other, then the operators $\chi_{i}g_{N}(D_{x})\chi_{j}$, and hence  $R_{n}$, will also be small.
\end{remark}
\begin{corollary}\label{cor-stup}
 Let $\un B$ be as in Prop. \ref{stup} and $\ti h_{i}\in \coinf(T^{*}\rr^{d})$ with $\pi_{x}\supp \ti h_{i}\cap \pi_{x}\supp \ti h_{j}= \emptyset$ for $i\neq j$, where $\pi_{x}: T^{*}\rr^{d}\to \rr^{d}$ is the projection on the configuration space. Let $R_{n}(t)$ be as in (\ref{R-n-definition}) with  the operators $h_{i}$ replaced with $\ti h_{i,t}^{\w}$. Then for any $\Delta\Subset\rr^{1+d}$ one has:
 \[
\| R_n(t)\one_{\Delta}(U)\|_{B(\hil)}\in O(t^{-\infty}).
\]
\end{corollary}
\proof We choose functions $\ti\chi_{i}\in \coinf(\rr^{d})$ such that $0\leq \ti\chi_{i}\leq 1$, $\ti\chi_{i}(x)=1$ near $\pi_{x}\supp \ti h_{i}$ and $\supp \ti\chi_{i}$ pairwise disjoint.  We set $\ti\chi_{i,t}(x):=\ti\chi_i(x/t)$ and denote the corresponding operators on $B(L^2(\real^d))$ by the same symbol.
 We apply Prop. \ref{stup} to $h_{i}= \ti h_{i,t}^{\w}$ and $\chi_{i}= \ti\chi_{i,t}$. By Prop.~\ref{pseudodifferential} (\ref{supports}), 
$\|\ti h_{i,t}^{\w}(1-\ti\chi_{i,t})\|+ \| (1-\ti\chi_{i,t})\ti h_{i, t}^{\w}\|\in O(t^{-\infty})$. Similarly we can estimate the operator norm of 
$\ti\chi_{i, t}g_{N}(D_x)\ti\chi_{j,t}$ by its Hilbert-Schmidt norm which equals
\[
\bigg(\int \ti\chi_{i,t}^{2}(x)\langle x- y\rangle^{-2N} \ti\chi_{j,t}^{2}(y)\, dxdy\bigg)^{\12}\in O(t^{d-N}).
\]
Since $N$ is arbitrary we obtain the lemma.
\qed

The following lemma is similar to \Proposition~\ref{stup}. Its proof is given in Appendix~\ref{B-2}.
\begin{lemma}\label{tazu}
 Let $B_{1}, B_{2}\in \mcL_{0}$, $h_{1}\in B(L^{2}(\rr^{d}))$ and $g_{2}\in L^{2}(\rr^{d})$.  Let us fix measurable functions $\chi_{i}: \real^{d}\to \real$, $i=1,2$,  with $0\leq \chi_{i}\leq 1$ and still denote by $\chi_{i}\in B(L^{2}(\real^{d}))$ the operator of multiplication by $ \chi_{i}$.  
 Then for any $\Delta\Subset \rr^{1+d}$, $N\in \nn$ there exists $C_{N}(\Delta, B_{1}, B_{2})$ such that:
 \beq \bes \label{zorglib}
 &\|[N_{B_{1}}(h_{1}), B_{2}^{*}(g_{2})]\one_{\Delta}(U)\|_{B(\cH)}\leq C_{N}(\Delta, B_{1}, B_{2})\times\\[2mm]
&\times\left(\|h_{1}(1-\chi_{1})\|_{B(L^{2}(\real^d))}\|g_{2}\|_{L^{2}(\real^d)}+ \| (1- \chi_{1})h_{1}\|_{B(L^{2}(\real^d))}\|g_{2}\|_{L^{2}(\real^d)}\right.\\[2mm]
&+\|h_{1}\|_{B(L^{2}(\real^d))}\|(1- \chi_{2})g_{2}\|_{L^{2}(\real^d) }\left.+ \|h_{1}\|_{B(L^{2}(\real^d))}\| g_{2}\|_{L^{2}(\real^d)}\| 
\chi_{1}g_{N}(D_{x})\chi_{2}\|_{B(L^{2}(\real^d))}\right).
\ees\eeq
\end{lemma}

\section{An intermediate convergence argument}\label{echi}\init

Theorem~\ref{Main-result-half} and Lemma~\ref{almost-KG-equation}  below
essentially reduce the proof of Theorem~\ref{Main-theorem} to an argument adapted from non-relativistic scattering theory,
which will be presented in Section~\ref{echa}.
The results of the
present section generalize to arbitrary $n$ the corresponding arguments from \cite{DG12}, where we studied the case of $n=2$ detectors.
The discussion in  Section~\ref{echa} will be very different from \cite{DG12}, however.

Let $B_i\in \mcL_{0}$, $h_{i}\in \cS(T^{*}\rr^{d})$, $i=1,\ldots, n$, and set  
\beq\label{def-de-Nb}
h_{i,t}(x,\xi):= h_i\left(\frac{x}{t},\xi \right), \ N_{B_i}(h_i^\w,t):= N_{B_{i,t}}(h_{i,t}^\w).
\eeq
We recall the notation $\hx=(x_1,\ldots,x_n)$ and $\tilde\omega(D_{\hx})= \omega(D_{x_{1}})+\cdots+\omega(D_{x_{n}})$,
where $\tilde\omega(D_{\hx})$ is an operator acting on $L^{2}(\rr^{nd})$.

\bet\label{Main-result-half} 
Let $\Delta\subset \rr^{1+d}$ be an open bounded set,  $B_{1},\ldots, B_{n}\in \mcL_{0}$ be s.t.   $\un B$  is $\Delta-$admissible
and let $h_{1},\ldots, h_{n}\in\cS(T^{*}\real^{d})$ be s.t. $\pi_{x}(\supp\, h_{i})$ are pairwise disjoint.  Let $\H(\hx,\hxi):=\prod_{i=1}^{n}h_{i}(x_{i}, \xi_{i})$ and 
$\H_t(\hx,\hxi):=\H(\hx/t,\hxi)$.

We  set for $\Psi\in \one_{\Delta}(U)\cH$:
\beq\label{defdeF-t}
F_{t}:= ( \langle \Omega|\otimes \one_{L^{2}(\real^{nd}) })\circ a_{\BB}\e^{-\i tH}\Psi\in L^{2}(\rr^{nd}), \,
\eeq
so that
\[
F_{t}(x_{1},\ldots, x_{n})= (\Omega| B_{1}(t, x_{1})\ldots B_{n}(t, x_{n})\Psi)_{\cH}, \quad (x_{1},\ldots, x_{n})\in \rr^{nd}.
\]
Assume that:
\beq\label{tiri}
F_{+}:= \lim_{t\to \infty}\e^{ \i t\tilde{\omega}(D_{\hx})}\H_{t}^\w F_{t}\hbox{ exists}.
\eeq
Then 
\beq\label{tara}
 \lim_{t\to \infty}N_{B_{1}}(h_{1}^\w, t)\ldots N_{B_{n}}(h_{n}^\w, t)\Psi
\eeq
exists and belongs to $\one_{\Delta}(U)\cH_{n}^{+}$.
\eet

\proof Applying Corollary \ref{cor-stup}, we get:
\[
\begin{array}{rl}
N_{B_{1}}(h_{1, t}^\w)\ldots N_{B_{n}}(h_{n,t}^\w)\one_{\Delta}(U)= a_{\ov{B}}^{*}\circ (\one_{\cH}\otimes \H^\w_{t})\circ a_{\BB}\one_{\Delta}(U)+ O(t^{-\infty}).
\end{array}
\]
By  the $\Delta-$admissibility of $\un B$ (more precisely, property (\ref{transfer-to-vacuum})) and Lemma \ref{toto.-1} we have:
\[
 a_{\BB}\one_{\Delta}(U)= ( \one_{\{0\}}(U)\otimes\one_{L^{2}(\real^{nd})})\circ a_{\BB}\one_{\Delta}(U)
 = ( |\Omega\rangle\langle \Omega|\otimes\one_{L^{2}(\real^{nd})  })\circ a_{\BB}\one_{\Delta}(U),
\]
using the spectrum condition (\ref{spec-ass}). Therefore we have:
\beq\bes\label{toto.e26}
\e^{\i tH}N_{B_{1}}(h_{1, t}^\w)\ldots N_{B_{n}}(h_{n,t}^\w)\e^{-\i tH}\Psi &= \e^{\i tH}a_{\BBc}^{*} (\Omega\otimes \H_{t}^\w F_{t}) + O(t^{-\infty})\\[2mm]
&=\e^{\i tH}a_{\BBc}^{*}(\Omega\otimes\e^{-\i t \tilde{\omega}(D_{\hx})}F_{+}) + o(t^{0}),
\ees\eeq
by (\ref{tiri}). We set
 \[
S_{t}: L^{2}(\rr^{nd})\ni F\mapsto  \ \e^{\i tH}a_{\BBc}^{*}(\Omega\otimes\e^{-\i t \tilde{\omega}(D_{\hx})}F) \in \cH.
\]
By Lemma \ref{toto.-1} the family $S_{t}$ is uniformly bounded in norm. Moreover if  $g_{1},\ldots, g_{n}$ are positive energy KG solutions with disjoint velocity supports  (defined in  Subsect. \ref{HR1}) and $f_{1},\ldots, f_{n}\in \cS(\real^d)$ are their initial data, then 
\[
S_{t}(f_{1}\otimes\cdots\otimes f_{n})= B^{*}_{1,t}(g_{1,t}) \ldots  B^{*}_{n, t}(g_{n,t})\Omega,
\]
where the Haag-Ruelle creation operators $B^{*}_{i, t}(g_{i,t})$ are defined in Subsect.~\ref{HR2}.  From Thm.~\ref{Haag-Ruelle} we know that $\lim_{t\to \infty}S_{t}(f_{1}\otimes\cdots\otimes f_{n})$ exists. By linearity and density arguments, using the uniform boundedness of $S_{t}$, we conclude that $\lim_{t\to\infty}S_{t}F$ exists for any $F\in L^{2}(\rr^{nd})$. In view of (\ref{toto.e26}) this implies the existence of the limit in (\ref{tara}).   
It is also clear from the approximation argument above  that this limit belongs to $\cH_{n}^{+}$.   Due to $\Delta-$admissibility of $\un B$ it belongs to the range of $\one_{\Delta}(U)$. \qed

\def\R{R}

The proof of the existence of the limit (\ref{tiri}) will be given in the next section.  The key input is
the fact that 
$F_{t}$ solves a Schr\"{o}dinger equation with Hamiltonian $\tilde{\omega}(D_{\hx})$ and a {\em source  term} $\R_{t}$ whose $L^{2}$ norm decreases very fast when $t\to +\infty$  outside of the collision planes $\{\, \hx\in\real^{nd}  \,:\, x_i=x_j\}$, $i\neq j$. This is the content of the
following lemma:

\begin{lemma}\label{almost-KG-equation}
 Let $F_{t}$ be defined in (\ref{defdeF-t}). Then:
 \ben
 \item $F_{t}$ is uniformly bounded in $L^{2}(\rr^{nd})$,
 \item $t\mapsto F_{t}\in L^{2}(\rr^{nd})$ is $C^{1}$ with 
 \[
\p_{t}F_{t}= -\i \tilde{\omega}(D_{\hx})F_{t}+\R_{t}, 
\]
where $\R_t$ satisfies the assumptions of Lemma~\ref{R-t-lemma} below.

 \een
\end{lemma}

\proof We have $F_{t}(x_{1},\ldots, x_{n})= (\Omega | B_{1}(t, x_{1}) \ldots B_{n}(t, x_{n})\Psi)_{\cH}$, which is uniformly bounded in $L^{2}(\rr^{nd})$  by Lemma \ref{toto.-1}.  We set $\dot B_i:=\pa_s B_i(s,0)_{|s=0}$
and note that
since $\Psi\in \cH_{\rm c}(U)$, the map $t\mapsto F_{t}\in L^{2}(\rr^{nd})$ is $C^{1}$ with:
\beq\bes
 \p_{t}F_{t}(\hx)&=\sum_{i=1}^n (\Omega |  B_{1}(t, x_{1})\ldots \dot B_i(t,x_i) \ldots B_{n}(t, x_{n})\Psi)_{\cH}\label{eq-derivation}\\  %[2mm]
&=\sum_{i=1}^n (\Omega | \dot B_i(t,x_i) B_{1}(t, x_{1})\ldots \check{i} \ldots B_{n}(t, x_{n})\Psi)_{\cH}+\R_{1,t}(\hx), 
\ees\eeq
where
\beq
\R_{1,t}(\hx)&=\sum_{i=2}^n\sum_{j=1}^{i-1} (\Omega |  B_{1}(t, x_{1})\ldots B_{j-1}(t,x_{j-1}) [B_j(t,x_j),\dot B_i(t,x_i)]\times\non\\
 &\pha{44444444444444444444444}\times B_{j+1}(t,x_{j+1})\ldots \check{i}\ldots B_{n}(t, x_{n})\Psi)_{\cH}.\non
\eeq
(For $j=1$ in the above sum  we set $B_{1}(t, x_{1})\ldots B_{j-1}(t,x_{j-1}):=I$  and
for $j=i-1$ we set  $B_{j+1}(t,x_{j+1})\ldots \check{i}\ldots B_{n}(t, x_{n})=B_{i+1}(t,x_{i+1})\ldots B_{n}(t, x_{n})$ which is to be understood as $I$ if $i=n$).
Note that $\dot B_i$ are again almost local by the definition of $\mcL_0$. Using almost locality of  $B_i$, $\dot B_{i}$,
we easily obtain that  $\R_{1,t}$ satisfies the assumptions of Lemma~\ref{R-t-lemma} below.

 There holds for any $\Phi\in \cH$:
\beq\bes\non
(\Omega| B_{i}(t,x_{i})\Phi)_{\cH}&=(\Omega | \one_{\{0\}}(U)B_{i}(t,x_{i})\Phi)_{\cH}=(\Omega| B_{i}(t,x_{i})\one_{H_{m}}(U)\Phi)_{\cH}\\[2mm]
&=(\Omega| B_{i}(x_{i})\e^{-\i t \omega(P)}\Phi)_{\cH}= \e^{- \i t \omega(D_{x_{i}})}(\Omega |B_{i}(x_{i})\Phi),
\ees\eeq
using (\ref{toto.e3}),  (\ref{transfer-to-hyperboloid}) and finally (\ref{toto.e0}). Differentiating this identity we obtain
\beq
(\Omega| \dot B_{i}(t,x_{i})\Phi)_{\cH}= -\i\omega(D_{x_{i}}) (\Omega| B_{i}(t,x_{i})\Phi)_{\cH}. \label{derivative-equation}
\eeq
We get from (\ref{eq-derivation}) and (\ref{derivative-equation}) that  
\beq\bes\non
\p_{t}F_{t}(\hx)&= 
-\sum_{i=1}^n\i\omega(D_{x_{i}}) (\Omega | B_i(t,x_i) B_{1}(t, x_{1})\ldots \check{i} \ldots B_{n}(t, x_{n})\Psi)_{\cH}+\R_{1,t}(\hx)\non\\
&=-\i \ti\om(D_{\hx})F_{t}(x_1,\ldots, x_n)+\R_{1,t}(\hx)+\R_{2,t}(\hx),
\ees\eeq
where
\begin{align}
\R_{2,t}(\hx):=&-\sum_{i=2}^n\sum_{j=1}^{i-1}\i\omega(D_{x_{i}})(\Omega |   B_{1}(t, x_{1})\ldots B_{j-1}(t,x_{j-1}) [B_i(t,x_i), B_j(t,x_j)]\times  \\
&\pha{4444444444444444444444444444}\times B_{j+1}(t,x_{j+1})  \ldots \check{i} \ldots B_{n}(t, x_{n})\Psi)_{\cH}.\non
\end{align}
To conclude the proof it suffices to show that $\R_{2,t}$ satisfies the assumptions of Lemma~\ref{R-t-lemma}.
To this end, we note that  for any $\Phi_1,\Phi_2\in \cH_{\rm c}(U)$ we can write
\beqa
\omega(D_{x_{i}})(\Phi_1|[B_i(t,x_i), B_j(t,x_j)]\Phi_2)=(\Phi_1|[C_i(t,x_i), B_j(t,x_j)]\Phi_2),
\eeqa
where,   by  Lemma \ref{newlemma} (3),
\[
C_{i}= (2\pi)^{-d/2}\int f(x)B_i(0,x) dx, \ f\in \cS(\rr^{d}), \ \widehat{f}(-p)\equiv \omega(p)\hbox{ near }\supp(\widehat B_{i}).
\]
Since $C_{i}$ are almost local,  we can argue as in the case of $R_{1,t}$ and  the proof is complete. \qed

\section{Scattering for  Schr\"odinger equations with  source terms}\label{echa}\init

In this section we give the proof of the existence of the limit 
\beq
F_{+}= \lim_{t\to +\infty}\e^{\i t \tilde{\omega}(D_{\hx})}\H_{t}^\w F_{t}, \label{F-plus}
\eeq
appearing in Thm. \ref{Main-result-half}.  The proof relies on the fact that $F_t$ satisfies a 
Schr\"odinger equation with a source term $R_t$ as shown in Lemma~\ref{almost-KG-equation} above.
  To control  the influence of  $R_{t}$, we need the following fact. 
 
\bel\label{R-t-lemma} Let $ \R_{t}  \in L^2(\real^{nd})$  satisfy
\beq\bes\non
i)\ \ &\| \R_{t}\|_{L^{2}(\real^{nd})}\in O(t^{N_{0}}), \ N_{0}\in \nn,\\ 
ii)\ \ &|\R_{t}(\un x)|\leq C_{N}\sum_{i\neq j}\langle x_{i}-x_{j}\rangle^{-N},  \ \forall \ N\in \nn, \hbox{ uniformly in }t.\\[-3mm]
\ees\eeq
Let $\H\in \cS(T^{*}\real^{nd})$ be such that 
\beq\label{tazi}
\pi_{\un x}(\supp\, \H)\cap \DD_{0}= \emptyset,
\eeq
where $\DD_{0}\subset \rr^{nd}$ is defined in  (\ref{diagonal}). Then
\beqa
\|\H_t^{\w} \R_t\|_{L^2(\real^{nd})}=O(t^{-\infty}).
\eeqa 
\eel
\begin{remark} 
Note that symbols which are admissible (in the sense of \Definition \ref{support-admissible}) satisfy (\ref{tazi}).
\end{remark}
\proof Let $K_{t}(\un x, \un y)= \widecheck{\H}((\un x+ \un y)/2t, \un x-\un y)$ be the distributional kernel of $\H_{t}^{\w}$. Let $\de>0$,  $\chi\in \coinf(\rr^{nd})$ with $\chi\equiv 1$ near $0$ and set $K_{1,t}(\un x,\un y)= K_{t}(\un x, \un y)\chi((\un x-\un y)/\delta t)$, $K_{2,t}= K_{t}- K_{1,t}$. Since $\widecheck{\H}\in \cS(\rr^{nd}\times \rr^{nd})$ we easily see that $\|K_{2,t}\|_{L^{2}(\rr^{2nd})}\in O(t^{-\infty})$. Still denoting by $K_{2,t}$ the operator with kernel $K_{2,t}$, we  deduce from {\it i)} that  $\|K_{2,t}\R_{t}\|_{L^{2}(\real^{nd})}\in O(t^{-\infty})$. On the other hand we have using {\it ii)} and (\ref{tazi}):
\beq\bes\non
&\|K_{1,t}\R_{t}\|^{2}_{L^{2}(\real^{nd})}\\[2mm]
& \leq C_{N}t^{3nd}\sum_{i\neq j}\int|\int |\widecheck{\H}((\un x+ \un y)/2, t(\un x- \un y)|\chi((\un x- \un y)/\delta)\langle t(y_{i}- y_{j} )\rangle^{-N}d\un y|^{2}d\un x.
\ees\eeq
 For $\delta\ll \epsilon$  the integrand is supported in $\{|y_{i}- y_{j}|\geq \epsilon/2\}$, hence the integral is $O(t^{3nd-2N})$.
This concludes the proof. \qed\\
%%%%%%%%%%%%%%%%%%%%%%%%%%%%%%%%%%%%%%%%%%%%%%%%%%%%%%%%%%%%%%%%%%%%%%%%%%%%%
The main ingredient of the proof of existence of the limit in (\ref{F-plus}) is  a  novel propagation
estimate established in Prop.~\ref{novel-propagation} below. As a preparation we recall that $F_{t}$ is defined 
in (\ref{defdeF-t}) and note that
for  $M\in \cS(T^{*}\real^{nd})$ and $M_{t}(\un x, \un \xi)= M(\un x/t, \un \xi)$ we have
by \Proposition~\ref{pseudodifferential}  (\ref{commutator}):
\beq 
\D M_{t}^{\w}&:= \p_{t}M_{t}^{\w}+ [\ti \omega(D_{\un x}), \i M^{\w}_{t}]= t^{-1}(dM)^{\w}_{t}+ O(t^{-2}),\hbox{ for}\label{taz}\\ 
 dM(\un x, \un \xi)&:= (\pa_t M_t-\{\ti\om(\un \xi), M_t\}) |_{t=1}= -(\un x-\nabla \ti \omega(\un \xi) )\cdot \nabla_{\un x}M(\un x ,\un \xi), 
\label{taz-one}
\eeq
where $\{\,\cdot\,,\,\cdot\,\}$ is the Poisson bracket. In the remaining part of this section we set $\|\,\cdot\,\|:=\|\,\cdot\,\|_{L^2(\real^{nd})}$.
%%%%%%%%%%%%%%%%%%%%%%%%%%%%%%%%%%%%%%%%%%%%%%%%%%%%%%%%%%
\begin{proposition}
 \label{novel-propagation}
Let  $\H\in \coinf(T^{*}\rr^{nd}; \rr^{nd})$  be admissible in the sense of \Definition~\ref{support-admissible}. Then
\beq  \bes   \label{propagation-est}
&\int_{1}^{+\infty} \|\left(\un x/t- \nabla \ti \omega(D_{\un x})\right)\cdot \H_t^\w F_t\|^2 \frac{dt}{t}<\infty,\\ 
& \int_{1}^{+\infty} \|\left(\un x/t- \nabla \ti \omega(D_{\un x})\right)\cdot \H_t^\w \e^{-\i t\ti\om(D_{\hx})}u\|^2\frac{dt}{t} \leq c\|u\|^2, \quad u\in L^2(\real^{nd}). 
\ees
\eeq
\end{proposition}
\proof 
Let $K$ and $\epsilon$ be as in \Definition~\ref{support-admissible}
and choose $0<\la\ll \epsilon$. Let $\chi\in C^{\infty}(\rr)$ with $\chi(s)= 1$ for  $ s\leq -1$, $\chi(s)=0$ for $s\geq (1+\la)^2$, $\chi'(s)\leq 0$ for $s\in\real$ and  $\chi'(s)\leq -\12$ for $0\leq s\leq 1$. Let  $\chi_{1}\in \coinf(\rr^{nd})$ be equal to $1$ on $K$ and
vanish outside of $K+\tiball(0,\la)$. We set
\[
M(\un x, \un \xi):=\chi(\epsilon^{-2}|\un x- \nabla \ti \omega(\un \xi)|^{2}) \chi_{1}(\nabla \ti \omega(\un \xi)).
\]
 By Def.~\ref{support-admissible},  $K+\tiball(0,\la)\subset \tiball(0,1)$, hence $ M\in \cS(T^{*}\rr^{nd})$  and we obtain from (\ref{taz-one})
\[
dM(\un x, \un \xi)=-\epsilon^{-2} \left|\un x - \nabla \ti \omega(\un \xi)\right|^{2}\chi'(\epsilon^{-2}|\un x- \nabla \ti \omega(\un \xi)|^{2}) \chi_{1}(\nabla \ti \omega(\un \xi)).
\]
 Making use of the properties of  $\chi$ and $\chi_1$ we obtain for some $c>0$: 
\beq
&dM(\un x, \un \xi)\geq c \one_{\ball(0,\epsilon)}(\un x-\nabla \ti \omega(\un \xi))\one_{K}(\nabla \ti \omega(\un \xi)) \left|\un x - \nabla \ti \omega(\un \xi)\right|^{2}, \label{lower-bound} \\ 
&\pi_{x}(\supp M)\subset \rr^{nd}\backslash \DD_{0}. \label{diagonal-avoidance}
\eeq
 Relation~(\ref{diagonal-avoidance}) holds for $\la\ll \epsilon$ and follows from the facts that for $\un y\in K+\tiball(0,\la)$ we have $|y_i-y_j|\geq 2(\epsilon-\la)$ for $i\neq j$ and
$|\un z|\leq \epsilon (1+\la)$ implies $|z_i|+|z_j|\leq \sqrt{2}\epsilon (1+\la)$.

Relation~(\ref{diagonal-avoidance})   and  Lemma \ref{R-t-lemma} imply that $\|M_{t}^{\w}R_{t}\|$, $\|(M_{t}^{\w })^{*}R_{t}\|$ belong to $L^{1}(\rr, dt)$.  Properties~(\ref{lower-bound}), (\ref{taz}) give
\[
 \D M_{t}^{\w}\geq \frac{c}{t}\left (\un x/t-\nabla\ti \omega(D_{\un x})\right)\cdot \H_{t}^{\w *}\H_{t}^{\w}\cdot  \left(\un x/t-\nabla\ti \omega(D_{\un x})\right)+ O(t^{-2}),
\]
using  \Proposition~\ref{pseudodifferential} (\ref{positivity}), (\ref{product}), (\ref{polynomials}). Applying Lemma \ref{A1} we obtain the first statement of (\ref{propagation-est}). The proof of the second is similar. \qed

%%%%%%%%%%%%%%%%%%%%%%%%%%%%%%%%%
\begin{theoreme}\label{Klein-Gordon-approximation}
 Let $F_{t}$,  be defined in (\ref{defdeF-t}) and $\H\in C_0^{\infty}(\real^{nd})$ be admissible in the sense of \Definition~\ref{support-admissible}. 
Then the limit
 \[
F_{+}= \lim_{t\to +\infty}\e^{\i t \tilde{\omega}(D_{\hx})}\H_{t}^\w F_{t}\hbox{ exists}.
\]
\end{theoreme}
\proof  
We proceed similarly as in the proof of \cite[Prop. 4.4.5]{DG97}. We apply (\ref{taz-one}) to
\beqa
M(\hx,\hxi):= \H(\hx,\hxi)- \left(\hx- \nabla \tilde{\omega}(\hxi) \right)\cdot (\nabla_{\hx} \H)(\hx,\hxi),\label{M-definition}
\eeqa
which yields
\beq
dM(\un x, \un \xi)= \left(\un x- \nabla\tilde{\omega}(\un \xi)\right)\cdot \nabla^{2}_{\un x}\H(\un x, \un \xi) \cdot \left(\un x- \nabla\tilde{\omega}(\un \xi)\right). \label{dM-first}
\eeq
By the admissibility of $\H$,  we can find  $\H_1\in C_0^{\infty}(T^*\real^{nd}; \real^{nd})$ which is also admissible s.t.
\beq
dM(\un x, \un \xi)= \left(\hx- \nabla \tilde{\omega}(\hxi) \right)\cdot \H_{1}(\un x, \un \xi)\nabla^{2}_{\un x}\H(\un x, \un \xi)\H_{1}(\un x, \un \xi)\cdot  \left(\hx- \nabla \tilde{\omega}(\hxi)\right).\label{dM-second}
\eeq
By pseudodifferential calculus (see \Proposition~\ref{pseudodifferential} (\ref{polynomials}),(\ref{product})) and (\ref{taz})
one has for $u\in L^{2}(\real^{nd})$:
\beq\label{Heisenberg-bound}
|(u| \D M_{t}^{\rm w}F_{t})|\leq \frac{c}{t}\|\left(\un x/t- \nabla \ti \omega(D_{\un x})\right)\cdot \H_{1,t}^\w u\|\,  \|\left(\un x/t- \nabla \ti \omega(D_{\un x})\right)\cdot \H_{1,t}^\w F_{t}\|+ O(t^{-2})\|u\|.
\eeq
  Since $M$ satisfies the assumptions imposed on $\H$ in Lemma~\ref{R-t-lemma}, we  obtain
\beqa
\| M_t^\w \R_{t}\|, \| (M_t^\w)^*\R_{t}\| \in O(t^{-\infty}). \label{rest-terms}
\eeqa
Making use of (\ref{Heisenberg-bound}),  (\ref{rest-terms}), \Proposition~\ref{novel-propagation} and Lemma~\ref{A2}, we get that
\[
\lim_{t\to +\infty}\e^{\i t \tom(D_{\hx})}M_tF_{t}\hbox{ exists}.
\]
To conclude the proof, it suffices to verify that $M_t-\H_t$  does not
contribute to the above limit.    For admissible $\H_2\in C_0^{\infty}(T^*\real^{nd}; \real^{nd})$, we set 
\[
N(\un x, \un \xi):= \left(\un x- \nabla \ti \omega(\un \xi)\right)\cdot \H_2(\un x, \un \xi).
\]
We have to show that
\beqa
\lim_{t\to\infty}\|N_t^\w F_t\|=0. \label{auxiliary-zero-limit}
\eeqa
 By \Proposition~\ref{novel-propagation} the limit must be zero if it exists. 
To prove the existence of the limit, we first note that by pseudodifferential calculus
\[
(N_{t}^{\w})^{*}N_{t}^{\w}= (|N|^{2}_t)^{\w}+ O(t^{-1}).
\]
Next, making use of relation~(\ref{taz-one}), we  obtain
\beq\bes
d |N|^{2}(\un x, \un \xi)&= ( \un x- \nabla \ti \omega(\un \xi))\cdot N_1(\un x,\un \xi) \cdot (\un x- \nabla \ti \omega(\un \xi) ), \hbox{ for }\label{dN-square}\\
N_{1, i,j}(\un x,\un \xi)&:= -( \un x-   \nabla \ti \omega(\un \xi))\cdot \nabla_{\un x}( \H_{2,i}^* \H_{2,j})(\un x,\un \xi)-2( \H_{2,i}^* \H_{2,j})(\un x,\un \xi),
\quad i,j=1,\ldots,nd.
\ees\eeq
Since $|N|^2$  is  admissible, we obtain by Lemma~\ref{R-t-lemma}:
\beq
\| (|N|^2_t)^\w\R_{t}\|\in O(t^{-\infty}).
\eeq
Exploiting the admissibility of $N_1$, we can rewrite (\ref{dN-square}) as in (\ref{dM-first}), (\ref{dM-second}) above and conclude the
existence of the limit (\ref{auxiliary-zero-limit})   from (\ref{taz}), \Proposition~\ref{novel-propagation} and Lemma \ref{A3}. \qed

\section{Haag-Ruelle scattering theory}\init\label{Haag-Ruelle-appendix}

In this section we collect some basic facts concerning the Haag-Ruelle scattering theory, which
we need for the proof of Theorem~\ref{Weak-asymptotic-completeness}. For the reader's convenience
we give a self-contained  presentation of this classical topic in the setting of the present paper.
In the special case of two-body scattering we presented a similar discussion  in \cite{DG12}.

\subsection{Positive energy solutions of the Klein-Gordon equation}\label{HR1}
\begin{definition} \label{KG-definition}
 Let  $f\in\cS(\rr^{d})$ be  such that $\widehat{f}$ has compact support. The function 
\[
g(t,x)= g_{t}(x)\hbox{ for }g_{t}= \e^{-\i t \omega(D_{x})}f,
\]
which solves $( \p_{t}^{2}- \Delta_{x}) g + m^{2}g=0$, 
will be called a  {\em positive energy KG solution}.
\end{definition}
%%%%%%%%%%%%%%%%%%%%%%%%%%%%%%%%%%%%%%%%%%%%%%%%%%%
 The following property of positive energy KG solutions is proven  in \cite{RS3}:
%%%%%%%%%%%%%%%%%%%%%%%%%%%%%%%%%%%%%%%%%%%%%%%%%%%
\bep \label{toto.20}
Let $\chi_{1}, \chi_{2}\in C^{\infty}(\rr^{d})$ be bounded with all derivatives and having disjoint supports. Let $f\in\cS(\rr^{d})$
be s.t. $\widehat f$ has compact support. Then
\[
\|\chi_{1}\left(\frac{x}{t}\right)\e^{- \i t \omega(D_{x})}\chi_{2}(\nabla \omega(D_{x}))f\|_{L^2(\real^d)} \in O(t^{-\infty}).
\]
\eep
%%%%%%%%%%%%%%%%%%%%
%%%%%%%%%%%%%%%%%%%%%%%%%%%%%%%%%%%%%%%%%%%%%%%%%%%%%%%%%%%%
We recall the  notion of  {\em velocity support} which will be useful later on.
\begin{definition}\label{def-de-vel}
 Let $\Delta\Subset H_{m}$. We set
 \[
{\rm Vel}(\Delta):= \{\nabla \omega(p)\ : \ p\in \rr^{d}, \ (\omega(p), p)\in \Delta\}.
\]
\end{definition}
\noindent It is  clear that disjointness of $\Delta_{1}$ and $\Delta_{2}$ entails that ${\rm Vel}(\Delta_{1})$ 
and ${\rm Vel}(\Delta_{2})$ are also disjoint. In view of  Prop.~\ref{toto.20} and of the fact that 
$\supp\,\widehat{g}\subset H_{m}$, we can call
 ${\rm Vel}(\supp \widehat{g})= \{\nabla \omega(p) \, : \,   p\in \supp \widehat{f}\}$
the {\em velocity support} of   a positive energy KG solution $g$ with initial data $f$.

\subsection{Haag-Ruelle scattering theory}\label{HR2}
Let $B\in  \mcL_{0}$ satisfy (\ref{transfer-to-hyperboloid}), that is $-\supp(\widehat B)\cap \Sp\, U\subset H_{m}$,
and let  $g$ be a positive energy KG solution. The {\em Haag-Ruelle creation operator} 
is given by 
\[
 B_{t}^{*}(g_{t})= \int g(t,x)B^{*}(t,x)dx,\quad t\in \real,
\]
which is well defined since $\e^{- \i t \omega(D_{x})}$ preserves $\cS(\rr^{d})$. The following lemma is
elementary, except for part (2) which relies on Lemma~\ref{existence-lemma}. We refer to \cite{DG12} for a  proof.

\begin{lemma}\label{toto.21}
 The following properties hold:
 \ben
 \item $B_{t}^{*}(g_t)\Omega= B^{*}(f)\Omega= (2\pi)^{d/2}\widehat{f}(P)B^{*} \Omega$, if $g_{t}= \e^{-\i t \omega(D_{x})}f$.
 \item Let $\Delta\Subset \rr^{1+d}$, $f\in L^{2}(\rr^{d})$. Then $\| B^{(*)}(f)\one_{\Delta}(U)\|\leq c_{\Delta, B}\| f\|_{L^{2}(\rr^{d})}$.
 \item $\p_{t}B_{t}^{*}(g_t)= \dot{B}^{*}_{t}(g_t)+ B_{t}^{*}(\dot{g}_t)$, where $\dot{B}= \p_{s}B(s, 0)_{\mid s=0}\in \mcL_{0}$ and $\dot{g}= \p_{t}g$ is a positive energy KG solution with the same velocity support as $g$.
 \een
\end{lemma}

The following  result  is known as  the  Haag-Ruelle theorem \cite{Ha58, Ru62}. 
In Appendix~\ref{H-R}  we give an elementary proof which uses ideas from \cite{He65, BF82, Ar99, Dy05} and exploits 
the bound  in Lemma \ref{toto.21} (2).
%%%%%%%%%%%%%%%%%%%%%%%%%%%%%%%%%%%%%%%%%%%%%%%%%%%%%%%%%%%%%
\bet\label{Haag-Ruelle} Let $B_1,\ldots, B_{n}\in \mcL_{0}$   satisfy (\ref{transfer-to-hyperboloid}). Let $g_1,\ldots,g_n$ be
positive energy KG solutions with disjoint velocity supports. Then: 
\ben
\item  There exists the  $n${\em -particle scattering state} given by
\beqa
\Psi^+=\lim_{t\to\infty} B^{*}_{1,t}(g_{1,t})\ldots B^{*}_{n,t}(g_{n,t})\Om. \label{scattering-state}
\eeqa
\item The state $\Psi^+$ depends only on the single-particle vectors $\Psi_i=B^{*}_{i,t}(g_{i,t})\Om$,  and therefore we can write $\Psi^+=\Psi_1\timeso\cdots\timeso\Psi_n$.  Given two such vectors $\Psi^+$ and $\tilde \Psi^+$ one has:
\beq
(\tilde \Psi^+|\Psi^+)&=\sum_{\si\in S_n}( \tilde \Psi_1|  \Psi_{\si_1})\ldots ( \tilde \Psi_n|  \Psi_{\si_n}),      \label{scalar-product}\\
U(t,x)(\Psi_1\timeso\cdots \timeso\Psi_n)&=(U(t,x)\Psi_1)\timeso\cdots\timeso (U(t,x)\Psi_n), \quad (t,x)\in \rr^{1+d}, \label{energy-factorization-relation}
\eeq
where $S_n$ is the set of permutations of an $n$-element set.
\een
\eet
%%%%%%%%%%%%%
Let us now explain how to obtain the (outgoing) $n${\em -particle wave operator}: 
 Let  
\[
\cH_{m}:= \one_{H_{m}}(U)\cH,
\]
be the space of {\em one-particle states}.  
For $\Psi_{1}, \ldots, \Psi_{n}\in \cH_m$ we set
\[
 \Psi_{1}{\otimes}_{\rm s}\cdots {\otimes}_{\rm s} \Psi_{n}:= \frac{1}{\sqrt{n!}} \sum_{\si\in S_n}  \Psi_{\si_1}\otimes\cdots\otimes \Psi_{\si_n}
\in \cH_m^{ \otimes_{\rm s} n }.
\]
%%%%%%%%%%%%%%%%%%%%%%%%%%%%%%%%%%%%%%%%%%%%%%%%%%%
\begin{proposition}
 \label{def-de-wave} For any $n\geq 1$
 there exists a unique isometry
 \[
W_{n}^{+}: \ \cH_{m}^{\otimes_{\rm s} n }\to \cH
\] 
with the following properties:
\ben
\item If $\Psi_{1},\ldots,\Psi_{n}$ are as in Thm. \ref{Haag-Ruelle}, then $W_{n}^{+}(\Psi_{1}{\otimes}_{\rm s}\cdots {\otimes}_{\rm s} \Psi_{n})= 
\Psi_{1}\timeso\cdots\timeso \Psi_{n}$. \vspace{1mm}
\item 
$U(t,x) \circ W_{n}^{+}= W_{n}^{+} \circ (U_m(t,x)\otimes\cdots \otimes U_m(t,x)), \ (t,x)\in \rr^{1+d}$, 
where we  denote by $U_m(t,x)$ the restriction of $U(t,x)$ to $\cH_{m}$.
\een
\end{proposition}
%%%%%%%%%%%%%%%%%%%%%%%%%%%%%%%%%%%%%%%%%%%%
\begin{definition}\label{main-Haag-Ruelle-concepts} Let $W_{n}^{+}$, $n\geq 1$, be the isometries defined in \Proposition~\ref{def-de-wave}
and let us define $W_{0}^{+}: \complex\Om\to \hil$ by $W_{0}^{+}\Om=\Om$. Let $\Ga(\cH_{m})$ be the symmetric Fock space over
$\cH_{m}$ and let $W^+: \Ga(\cH_{m})\to \hil$ be the isometry given by    $W^+:=\bigoplus_{n\geq 0}W_{n}^{+}$. 
\ben
\item The map $W_{n}^{+}: \      \cH_{m}^{\otimes_{\rm s} n }   \to \cH$  is called the  (outgoing) $n${\em -particle wave operator}. 
\item The map $W^+: \Ga(\cH_{m})\to \hil$ is called the   (outgoing) {\em  wave operator}.
\item  The range of $W_{n}^{+}$ is denoted by $\cH_{n}^{+}$ and called the subspace of $n${\em-particle scattering states}.
\item The range of $W^{+}$ is denoted by  $\cH^{+}$   and called the subspace of {\em scattering states}.
 \een
\end{definition}
%%%%%%%%%%%%%%%%%%%%%%%%%%%%%%%%%%%%%%%%%%%%%%%
\noindent {\em Proof of Prop. \ref{def-de-wave}.} Let  ${\mathcal F}\subset  \cH_{m}^{\otimes_{\rm s} n }$ be the subspace spanned by  vectors $\Psi_{1} \otimes_{\rm s} \cdots \otimes_{\rm s}   \Psi_{n}$ for $\Psi_{1},\ldots,\Psi_{n}$ as in Thm. \ref{Haag-Ruelle}.  Due to (\ref{scalar-product})
 there exists a unique isometry $W_{n}^{+}: {\mathcal F}\to \cH$ such that 
 \[
W_{n}^{+}(\Psi_{1}{\otimes}_{\rm s} \ldots  {\otimes}_{\rm s}  \Psi_{n})= \Psi_{1}\timeso\ldots\timeso \Psi_{n},
\]
for all $\Psi_{1},\ldots, \Psi_{n}$ as in the theorem. Also,  by (\ref{energy-factorization-relation}), 
\beq
U(t,x)\circ W_{n}^{+}= W_{n}^{+} \circ (U_m(t,x)\otimes\cdots\otimes U_m(t,x)).  
\eeq
Thus  it suffices to prove that the closure of ${\mathcal F}$ is $\cH_{m}^{\otimes_{\rm s} n } $. 

Let $(H_{i}, P_{i})$, $i=1,\ldots,n$, be the generators of the groups of unitaries 
\beqa
(t,x)\mapsto \underbrace{\one\otimes\cdots\otimes\one}_{i-1}\otimes U_m(t,x) \otimes\underbrace{\one\otimes\cdots \otimes\one}_{n-i}, 
\eeqa
acting on $\cH_{m}^{\otimes n }$. We note that the joint spectral measure of $(\tilde{H}, \tilde{P}):= ((H_{1}, P_{1}),\ldots, (H_{n}, P_{n}))$ is supported by $H_{m}^{\times n}$.

 Let $B\in\mcL_{0}$ satisfy (\ref{transfer-to-hyperboloid}) and $g$ be a positive energy KG solution.
Then, due to Lemma~\ref{toto.21}~(1) and  the cyclicity of the vacuum, the set of vectors $B^{*}_{t}(g_t)\Om$ 
 is dense in $\cH_{m}$.
 Also, for $\Delta\Subset H_{m}$, the set of such vectors  with $g$ having the velocity support included in ${\rm Vel}(\Delta)$ is dense in $\one_{\Delta}(U)\cH_{m}$. Thus the closure of ${\mathcal F}$ in $\cH_{m}^{\otimes_{\rm s} n }$ equals  
\[
 {\mathcal F}^{\rm cl}= \Theta_{\rm s} \circ \one_{(H_{m}^{\times n})\backslash D}(\tilde{H}, \tilde{P})(\cH_{m}^{\otimes n}),
 \]
 where $\Theta_{\rm s}: \cH_{m}^{\otimes n} \to \cH_{m}^{\otimes_{\rm s} n }$ is the orthogonal projection, and 
\beqa
D:=\{\, \un p\in H_{m}^{\times n}\,:\, p_i=p_j\textrm{ for some } i\neq j \}.  
\eeqa 
By \cite[Prop. 2.2]{BF82}, the spectral measure of  the restriction of $(H, P)$ to $\cH_{m}$ is absolutely continuous w.r.t. the Lorentz invariant measure on $H_{m}$. Hence  $\one_{D}(\tilde{H}, \tilde{P})=0$, which completes the proof. \qed

\section{Proof of \Theorem  \ref{Weak-asymptotic-completeness} }\init\label{blit}

We recall that the notation $N_B(h_1)$, $N_{B}(h_2^\w, t)$ was introduced in Def. \ref{toto.2}
and in (\ref{def-de-Nb}), respectively, for $B\in \mcL_{0}$, $h_1\in B(L^2(\real^d))$ and $h_2\in \cS(T^*\real^d)$.

\begin{proposition}\label{C1}
 Let $i= 1,\ldots, n$. Let   $\Delta_{i}\Subset H_{m}$ be disjoint sets, $B_{i}\in \mcL_{0}$ and  
$\supp(\widehat B_{i})$ be disjoint sets. Assume moreover that:
 \beq
&-\supp(\widehat B_{i})\cap \Sp\, U\subset \Delta_{i},\label{e.c1a}\\ 
&(\Delta_{i}+ \supp(\widehat B_{i}))\cap \Sp (U)\subset \{0\}.  \label{e.c1b}
\eeq
 Let $h_{i}\in C_0^{\infty}(  T^* \rr^{d})$ such that  
\beq\bes
\label{taza}
(1) \ \ & h_i(y,(\nabla\om)^{-1}(y))=1, \  y\in  {\rm Vel}(\Delta_{i}),\\ 
(2)\ \ &\pi_{x}\supp\, h_i\cap\pi_{x}\supp \, h_j=\emptyset, \ \forall\   i\neq j.
\ees \eeq
Then, for $\Psi_{i}\in \one_{\Delta_{i}}(U)\cH$,
\begin{equation}
\label{e.c4}
\lim_{t\to +\infty} N_{B_{1}}(h_{1}^\w, t) \ldots N_{B_{n}}(h_{n}^\w, t)W_{n}^{+}(\Psi_{1}\otimes_{\rm s}\cdots\otimes_{\rm s} \Psi_{n})
= W_{n}^{+}(N_{B_{1}}(\one)\Psi_{1}\otimes_{\rm s}\cdots\otimes_{\rm s} N_{B_{n}}(\one)\Psi_{n}).
\end{equation}
\end{proposition}
\begin{remark}
  Note that  $W_{n}^{+}(\Psi_{1}\otimes_{\rm s}\cdots \otimes_{\rm s}  \Psi_{n})$ belongs to $\cH_{\rm c}(U)$, and that $N_{B_{i}}(\one)\Psi_{i}$ belong to $\one_{\Delta_{i}}(U)\cH$, because of (\ref{e.c1a}), (\ref{e.c1b}).  Hence  all the expressions appearing in (\ref{e.c4}) are well defined.
 \end{remark}
 
\proof Due to the fact that   $N_{B_{i}}(\one)\Psi_{i}$ 
satisfy the assumption imposed on $\Psi_i$ in the proposition, it suffices to show that
\beqa
\lim_{t\to +\infty} N_{B_{i}}(h_{i}^\w, t)W_{n}^{+}(\Psi_{1}\otimes_{\rm s}\cdots\otimes_{\rm s} \Psi_{n})
= W_{n}^{+}(\Psi_{1}\otimes_{\rm s}\cdots    \otimes_{\rm s}N_{B_{i}}(\one)\Psi_{i}\otimes_{\rm s}\cdots\otimes_{\rm s} \Psi_{n})\label{iteration}
\eeqa
and then iterate this result making use of the bound $\sup_{t\in\real}\|N_{B_{i}}(h_{i}^\w, t) \one_{\Delta}(U)\|<\infty$ valid for any $\De\Subset\real^{1+d}$. By the same token,
it suffices  to assume that $\Psi_{j}= A^{*}_{j, t}(g_{j,t})\Omega$ for $A_{j}\in \mcL_{0}$ 
satisfying (\ref{e.c1a}) and $g_{j}$ a positive energy KG solution with the velocity support 
included in ${\rm Vel}(\Delta_{j})$, so that $\Psi_{j}= \one_{\Delta_{j}}(U)\Psi_{j}$. 
Similarly, since $N_{B_{i}}(\one)\Psi_{i}\in\one_{\Delta_{i}}(U)\cH$, we can find for any $0<\epsilon_{i}\ll 1$ operators $\tilde{A}_{i}\in \mcL_{0}$ and positive energy KG solutions $\tilde{g}_{i}$, satisfying  the same properties as $A_{i}, g_{i}$, such that  
\begin{equation}
\label{e.c11}
\| N_{B_{i}}(\one)\Psi_{i}- \tilde{A}_{i,t}^{*}(\tilde{g}_{i,t})\Omega\|\leq \epsilon_{i}, \  i=1,\ldots,n.
\end{equation}
We fix such $A_{j}, g_{j}$  and $\ti A_i, \ti g_i$ for future reference.

First, we  claim that for $B,\Delta, \Psi, h$  as in the proposition one has:
 \begin{equation}
\label{e.c0}
\lim_{t\to+\infty}N_{B}(h^\w, t)\Psi= N_{B}(\one)\Psi.
\end{equation}
In fact,  due to (\ref{e.c1a}), (\ref{e.c1b}) we have 
\beq\label{tarali}
B^{*}B\one_{\Delta}(U)=B^{*}|\Omega\rangle\langle \Omega|B\one_{\Delta}(U) =\one_{\Delta}(U)B^{*}B \one_{\Delta}(U).
\eeq 
Therefore,
\beq\bes\non
N_{B}(h^\w, t)\Psi&=\e^{\i tH}N_{B}(h^\w_{t})\e^{-\i tH}\Psi\\[2mm]
&=\e^{\i t\omega(P)} a_{B}^{*}\circ (\one_{\cH}\otimes h^\w_{t})\circ a_{B}\e^{- \i t \omega(P)}\Psi\\[2mm]
&=a_{B}^{*}\circ \e^{\i t \omega(P+ D_{x})}(\one_{\cH}\otimes h^\w_{t})\e^{-\i t \omega(P+D_{x})} \circ a_{B}\Psi\\[2mm]
&=a_{B}^{*}\circ (\one_{\cH}\otimes \e^{\i t \omega(D_{x})} h^\w_{t} \e^{-\i t \omega(D_{x})}  )\circ a_{B}\Psi, 
\ees\eeq
where we used (\ref{toto.e0}) and the fact that $a_{B}\Psi=(|\Om\ran \lan\Om|\otimes\one_{L^2(\real^d) })\circ a_{B}\Psi$. Making use of \Proposition~\ref{pseudodifferential}~(\ref{limit}), we get
\beqa
\slim_{t\to\infty}\e^{\i t \omega(D_{x})} h^\w_{t} \e^{-\i t \omega(D_{x})}=h(\nabla\om(D_x),D_x).
\eeqa
Thus we obtain
\begin{equation}
\lim_{t\to +\infty}N_{B}(h^\w, t)\Psi=a_{B}^{*}\circ (\one_{\cH} \otimes h(\nabla \omega(D_{x}),D_x))\circ a_{B}\Psi= a_{B}^{*}a_{B} h(\nabla \omega(P),P)\Psi,
\label{relevant-to-introduction}
\end{equation}
exploiting Lemma~\ref{toto.1} and once again (\ref{toto.e0}). By (\ref{taza}) (1)  we have $h(\nabla \omega(p),p)=1$ for $(\omega(p), p)\in \Delta$. Hence
$h(\nabla \omega(P),P)\Psi= \Psi$, which completes the proof of (\ref{e.c0}).

Next, we claim  that for $i\neq j$:
\begin{equation}
\label{e.c10}
\| [N_{B_{i}}(h_{i}^\w, t), A_{j,t}^{*}(g_{j,t})]\| \in O(t^{-\infty}).
\end{equation}
To show (\ref{e.c10}), we first note that
${\rm Vel}(\Delta_{j})\subset \pi_{x} \supp\, h_{j}$ by (\ref{taza}) (1). Hence $\pi_{x}\supp\, h_{i}$ and the velocity support of $g_{j}$ are disjoint by (\ref{taza}) (2).
 Let $\chi_{i}, \chi_{j}\in \coinf(\rr^{d})$ with  $0\leq \chi_{i}, \chi_{j}\leq 1$, $\supp \chi_{i}\cap\supp \chi_{j}= \emptyset$ and  $\chi_{i}\equiv 1$ near $\pi_{x}\supp\,h_{i}$, $\chi_{j}\equiv 1$ near the velocity support of $g_{j}$. 
We set $\chi_{i,t}(x):=\chi_i(x/t)$, $\chi_{j,t}(x):=\chi_j(x/t)$ and denote the corresponding operators on $L^2(\real^d)$ by the same symbols.
We recall that $g_{N}(\xi)$ is defined in (\ref{toto.e13bis}) and note that
\begin{equation}
\label{tazo}
\|(1-\chi_{i,t})h_{i,t}^{\w}\|_{B(L^{2}(\real^d) )}, \  
\| h_{i,t}^{\w}(1-\chi_{i,t})\|_{B(L^{2}(\real^d))},\ 
\| \chi_{i,t} g_{N}(D_{x})\chi_{j,t}\|_{ B(L^{2}(\real^d)) }\in O(t^{-\infty}), 
\end{equation}
where the expressions involving $h_{i,t}^{\w}$ are treated using 
Prop.~\ref{pseudodifferential} (\ref{supports}) and the expression with $g_N$ by
inspection of its kernel as in the proof of Corr.~\ref{cor-stup}.
By \Proposition~\ref{toto.20} we have:
\begin{equation}
\label{toza}
\|(1- \chi_{j,t})g_{j,t}\|_{L^{2}(\real^d)}\in O(t^{-\infty}).
\end{equation}
Then  (\ref{e.c10}) follows by applying Lemma \ref{tazu} for $B_{1}= B_{i}$, $B_{2}= A_{j}$, $h_{1}= h_{i,t}^{\rm w}$, $g_{2}= g_{j,t}$ and $\chi_{1}=\chi_{i,t}$, $\chi_{2}= \chi_{j,t}$. In fact the  quantities in the r.h.s. of (\ref{zorglib}) are $O(t^{-\infty})$ by (\ref{tazo}) and (\ref{toza}).

After these preparations we proceed to the proof of (\ref{iteration}). 
Using  (\ref{e.c10}), (\ref{e.c0}), we obtain:
\beq\bes\non
N_{B_{i}}(h_{i}^\w, t)(\Psi_{1}\timeso \cdots \timeso\Psi_{n})&=N_{B_{i}}(h_{i}^\w, t)A_{1,t}^{*}(g_{1,t}) \ldots  A_{n, t}^{*}(g_{n,t})\Omega+ o(t^{0})\\ 
&= A_{1,t}^{*}(g_{1,t})\ldots\check{i} \ldots A_{n, t}^{*}(g_{n,t})N_{B_{i}}(\one)\Psi_{i} + o(t^{0})\\
&=A_{1,t}^{*}(g_{1,t})\ldots\check{i} \ldots A_{n, t}^{*}(g_{n,t})  \tilde{A}_{i,t}^{*}(\tilde{g}_{i,t})\Omega + o(t^{0})+O(t^{0})\epsilon_{i}\\ 
&=\Psi_{1}\timeso \cdots \timeso\tilde{\Psi}_{i}\timeso \cdots\timeso\Psi_{n}+ o(t^{0})+O(t^{0})\epsilon_{i}\\ 
&=\Psi_{1}\timeso \cdots \timeso  N_{B_{i}}(\one)\Psi_{i}   \timeso \cdots\timeso\Psi_{n}+ o(t^{0})+O(t^{0})\epsilon_{i},
\ees\eeq
where  $\tilde{\Psi}_{i}:=  \tilde{A}_{i,t}^{*}(\tilde{g}_{i,t})\Omega$. Since $\epsilon_i>0$ is arbitrary, this concludes the proof of the proposition. \qed

\begin{lemma}
 \label{Delta-inclusions}
  Let $\Delta\subset G_{2m}$ be an open bounded set. Then
\beq
\one_{\Delta}(U)\cH_{n}^{+}= {\rm Span}\{W_{n}^{+}(\Psi_{1}\otimes_{\rm s}\cdots \otimes_{\rm s}\Psi_{n})\ : \ \Psi_{i}\in \one_{\Delta_{i}}(U)\cH, \ \Delta_{i}\Subset H_{m}, \ \Delta_{i}\cap \Delta_{j}=\emptyset,\ i\neq j,\non\\ \Delta_{1}+\cdots+ \Delta_{n}\subset \Delta  \}^{\rm cl}. \non
\eeq
\end{lemma}
\proof The statement follows immediately from Prop. \ref{def-de-wave} (2) and the absolute continuity of the spectral measure of $(H,P)$ restricted to $\cH_{m}$, recalled in its proof. \qed

\bel\label{O-lemma} Let $\De\subset G_{2m}$ be an open bounded set s.t. $(\ov{\De}-\ov{\De})\cap\Sp\,U= \{0\}$. 
Let $\De_1,\ldots,\De_n\Subset H_{m}$ be  disjoint and such that  $\De_1+\cdots+\De_n\subset \De$.  Then there exist 
$O_1,\ldots, O_n\subset\real^{1+d}$ which are disjoint open neighbourhoods of $\De_1,\ldots,\De_n$, respectively, s.t. for any 
 $K_1,\ldots, K_n\Subset\real^{1+d}$ satisfying  $-K_i\subset O_i$, $-K_i\cap \Sp\, U\subset \De_i$, $i=1,\ldots, n$, one has:
\beq
& (\ov{\De}+K_1+\cdots+K_n)\cap \Sp\, U\subset \{0\}, \label{transfer-to-vacuum-one}\\
& -(K_1+\cdots+K_n)\subset \De, \label{transfer-from-vacuum-one}\\
& (\De_i+K_i)\cap\Sp\,U\subset \{0\}. \label{smallness-of-transfers-one}
\eeq
\eel
\proof Assume that  $O_i\subset \De_i+\ball_1(0,\eps)$, where 
$\ball_1(0,\eps):=\{\, x\in \real^d\,:\, |x|<\eps\,\}$. 
To prove (\ref{transfer-to-vacuum-one}), we write
\beq
\ov{\De}+K_1+\cdots+K_n\subset \ov{\De}-(O_1+\cdots +O_n)&\subset \ov{\De}-(\De_1+\cdots+\De_n)+\ball_1(0,n\eps)\non\\
&\subset  \ov{\De}-\ov{\De}+\ball_1(0,n\eps).\non
\eeq
Since, by assumption,  $(\ov{\De}-\ov{\De})\cap\Sp\, U= \{0\}$  and $0$ is isolated in $\Sp\, U$, 
we obtain that $(\ov{\De}-\ov{\De}+\ball_1(0,n\eps))\cap\Sp\,U=\{0\}$ for  $\eps\ll 1$.
As for (\ref{transfer-from-vacuum-one}), we obtain that 
\beqa
-(K_1+\cdots+K_n)\subset O_1+\cdots+O_n\subset \De_1+\cdots+\De_n+\ball_1(0,n\eps)\subset \De,\non
\eeqa 
for $\eps\ll 1$  using that $\Delta_{i}$ are compact and $\De$ is open.
Finally we write:
\beqa
\De_i+K_i\subset O_i-O_i\subset \Delta_{i}- \Delta_{i}+ \ball_1(0, 2\eps).\non
\eeqa 
We note  that  a difference of two vectors from $H_m$ is  either $0$  or spacelike.  For $\eps\ll 1$ we obtain (\ref{smallness-of-transfers-one}). \qed

%%%%%%%%%%%%%%%%%%%%%%%%%%%%%%%%%%%%%%%%%%%%%%
\bel\label{subspace-equality-lemma} Let $\De\Subset H_m$  and $O\subset \real^{1+d}$ be a sufficiently small neighbourhood 
of $\De$. Then 
\[
\one_{\Delta}(U)\hil=\Span\{\, N_{B}(\one)\one_{\Delta}(U)\hil   \setbar B\in \mcL_0,\,  -\supp(\widehat B)\subset O, 
-\supp(\widehat B)\cap\Sp\,U\subset \De\, \}^{\cl}.
\]
\eel
%%%%%%%%%%%%%%%%%%%%%%%%%%%%%%%%%%%%%%%
\proof  A proof of this lemma, which is based on ideas from \cite[Thm.~3.5]{DT11},  can be found in \cite{DG12}. \qed

\noindent {\em Proof of Thm. \ref{Weak-asymptotic-completeness}. } By Thm.~\ref{Main-theorem}, it is enough  to verify that
\beqa
\one_{\Delta}(U)\cH_{n}^{+}\subset \Span\{  
\Ran\, \C_{n,\al}^+(\De)\setbar     \al\in J  \}^{\cl}.
\eeqa
In view of  Lemma~\ref{Delta-inclusions}, it suffices to show that for any  $\De_i\Subset H_m$, $i=1,\ldots,n$,  such that 
$\De_1+\cdots+\De_n\subset \De$ and $\De_i\cap \De_j=\emptyset$  for $i\neq j$ one has
\beqa
W_{n}^{+}\left(\one_{\Delta_{1}}(U)\cH\otimes_{\rm s}\cdots \otimes_{\rm s}\one_{\Delta_{n}}(U)\cH\right)\subset  \Span\{  \Ran\, \C_{n,\al}^+(\De)\setbar     \al\in J  \}^{\cl}. \label{small-inclusion}
\eeqa 
Let $O_i\subset \real^{1+d}$ be sufficiently small open neighbourhoods of $\De_i$
so that the assertions of Lemma~\ref{O-lemma} hold. Let $B_i\in \mcL_0$ be such that   $-\supp(\widehat B_{i})\subset O_i$, $-\supp(\widehat B_{i})\cap \Sp\, U\subset \De_i$.
By Lemma~\ref{O-lemma}, $B_i$ are $\De-$admissible in the sense of \Definition~\ref{delta-admissible} and satisfy the assumptions of 
Prop.~\ref{C1}. 
We also choose $h_i\in C_0^{\infty}(T^*\real^d)$ as in  Prop. \ref{C1} and s.t. $h_1\otimes\cdots \otimes h_n$ is admissible 	 
in the sense of \Definition~\ref{support-admissible}. For example one can choose  $h_i(x,\xi):=h_{0,i}(x)\chi(x-\nabla\om(\xi))$, where $h_{0,i}\in C_0^\infty(\real^d)$
are equal to one on ${\rm Vel}(\Delta_{i})$ and have disjoint supports contained in the unit ball. The function  
$\chi\in C_0^{\infty}(\real^d)$ satisfies $\chi(0)=1$ and 
is supported in a sufficiently small neighbourhood of zero, depending
on the supports of $h_{0,i}$.

Let $J_0$ be the set of pairs $(\un B, \un h)$ as specified above.   We get
\beq\bes\label{subspace-inclusion}
&\Span \{\C_{n,\alpha}^+(\De)\circ W_{n}^{+}\left(\one_{\Delta_{1}}(U)\cH\otimes_{\rm s}\cdots\otimes_{\rm s} \one_{\Delta_{n}}(U)\cH\right)\ : \ \alpha\in J_{0} \} \\
 &=\Span\{W_{n}^{+}\left(N_{B_{1}}(\one)\one_{\Delta_{1}}(U)\cH\otimes_{\rm s}\cdots \otimes_{\rm s} N_{B_{n}}(\one)\one_{\Delta_{n}}(U)\cH\right)\ : \ \alpha\in J_{0}\}\\ 
 &= W_{n}^{+}\left(\one_{\Delta_{1}}(U)\cH\otimes_{\rm s}\cdots   \otimes_{\rm s} \one_{\Delta_{n}}(U)\cH\right).
\ees\eeq
In the first step we  used  Prop. \ref{C1} and in the second  Lemma~\ref{subspace-equality-lemma}.
Since $J_0\subset J$, the subspace on the l.h.s. of (\ref{subspace-inclusion}) is included in the subspace on the r.h.s. of (\ref{small-inclusion}).
This concludes the proof. \qed

\appendix

\section{Propagation estimates for inhomogeneous evolution equations}\init\label{alo}
In this section, which appeared already in \cite{DG12}, we extend standard results on propagation estimates and existence of asymptotic observables to the case of an inhomogeneous evolution equation:
\[
\p_{t}u(t)= -\i H u(t)+ r(t).
\]
Let $\cH$ be a Hilbert space and $H$ a self-adjoint operator on $\cH$.  We choose a function 
\[
\rr^{+}\ni t\mapsto u(t)\in \cH,
\]
 such that
\beq\bes\label{abs.e1}
i) \ \ & \sup_{t\geq 0}\| u(t)\|<\infty,\\ 
ii) \ \ & u(t)\in C^{1}(\rr^{+}, \cH)\cap C^{0}(\rr^{+}, \Dom H),
\ees
\eeq
and define:
\[
r(t):= \p_{t}u(t)+ \i  H u(t).
\]
For a map  $\rr^{+}\ni t\mapsto M(t)\in B(\cH)$ we denote by $\D M(t)= \p_{t}M(t)+ [H, \i M(t)]$ the Heisenberg derivative of $M(t)$. We assume that  $[H, \i M(t)]$, defined first as a quadratic form on $\Dom H$, extends by continuity to a bounded operator.

The following three lemmas can be proved by modifying standard arguments, see e.g. \cite[Sect. B.4]{DG97}.
By $C_j(\,\cdot\,)$, $B(\,\cdot\,)$, $B_1(\,\cdot\,)$ we denote auxiliary functions from $\real^+$ to $B(\hil)$.
\begin{lemma}\label{A1}
 Let $\rr^{+}\ni t\mapsto M(t)\in B(\cH)$ be s.t.:
 \beq\bes\non
i) \ \ &\sup_{t\in \rr^{+}}\|M(t)\|<\infty,\quad  \|M(\cdot)r(\cdot)\|, \ \|M^{*}(\cdot)r(\cdot)\|\in L^{1}(\rr^{+},dt),\\[-2mm]
ii) \ \ &\D M(t)\geq B^{*}(t)B(t)- \sum_{j=1}^{n}C_{j}^{*}(t)C_{j}(t),\ \int_{\rr^{+}}\| C_{j}(t)u(t)\|^{2}dt<\infty.
\ees\eeq
Then 
\[
 \int_{\real^+} \| B(t)u(t)\|^{2}dt<\infty.
\]
\end{lemma}
\begin{lemma}\label{A3}
 Let $\rr^{+}\ni t\mapsto M(t)\in B(\cH)$ be s.t.:
 \beq\bes\non
i) \ \ &\sup_{t\in \rr^{+}}\|M(t)\|<\infty,\quad \|M(\cdot)r(\cdot)\|, \ \|M^{*}(\cdot)r(\cdot)\|\in L^{1}(\rr^{+},dt),\\[-2mm]
ii) \ \ &|(u_{1}|\D M(t)u_{2})|\leq\sum_{j=1}^{n}\|C_{j}(t)u_{1}\|\|C_{j}(t)u_{2}\|,\ u_{1}, u_{2}\in\cH, \\[-2mm]
\hbox{ with }&\int_{\rr^{+}}\| C_{j}(t)u(t)\|^{2}dt<\infty.
\ees\eeq
Then 
\[
\lim_{t\to +\infty}(u(t)| M(t)u(t))\hbox{ exists}.
\]
\end{lemma}
\begin{lemma}\label{A2}
Let $\rr^{+}\ni t\mapsto M(t)\in B(\cH)$ be s.t.:
\beq\bes\non
i) \ \ &\| M(\cdot)r(\cdot)\|\in L^{1}(\rr^{+},dt),\\ 
ii)\ \ &|(u_{1}| \D M(t)u(t))|\leq \| B_{1}(t)u_{1}\|\| B(t)u(t)\|,\hbox{ with}\\ 
iii)\ \ &\int_{\rr^{+}} \| B(t)u(t)\|^{2}dt <\infty, \ \int_{\rr^{+}}\| B_{1}(t)\e^{-\i tH}u_{1}\|^{2}dt\leq C \| u_{1}\|^{2}, \ u_{1}\in \cH.
\ees\eeq
Then 
\[
\lim_{t\to +\infty}\e^{\i tH}M(t)u(t)\hbox{ exists}.
\]
\end{lemma}
\section{Some technical proofs}\init\label{stup-proof}
In this section we give the proofs of \Proposition~\ref{stup}, Lemma \ref{tazu}, Prop.~\ref{pseudodifferential}
and Thm.~\ref{Haag-Ruelle}.

\subsection{Proof of \Proposition~\ref{stup}.}
\proof   We will prove the proposition by induction on $n$.  We   set  $\un x= (x_{1}, \dots , x_{n}), \un y= (y_{1}, \dots ,y_{n})\in \rr^{nd}$ and denote by $h_{i}(x_{i}, y_{i})$ the distributional kernel of $h_{i}$.   

 We will also  write $\tilde{h}_{i}= h_{i}\chi_{i}$ for $1\leq i\leq n-1$ and $\tilde{h}_{n}= \chi_{n}h_{n}$.\def\th{\tilde{h}} Note that 
\beq\bes \label{eapp.-1}
\th_i(x_{i}, y_{i})&= h_{i}(x_{i}, y_{i})\chi_{i}(y_{i}), \ 1\leq i\leq n-1, \\ 
\th_n(x_{n}, y_{n})&= \chi_{n}(x_{n})h_{n}(x_{n}, y_{n}).
\ees\eeq
  We will first estimate  the analog of $R_{n}$ with $h_{i}$ replaced with $\tilde{h}_{i}$, which will be denoted by $\tilde{R}_{n}$.
Note first that  since $B_{i}$ have compact energy-momentum transfers,  for any $\Delta\Subset \rr^{1+d}$ there exists $\Delta'\Subset \rr^{1+d}$ such that:
\[
\tilde{R}_{n}\one_{\Delta}(U)= \one_{\Delta'}(U)\tilde{R}_{n}\one_{\Delta}(U),
\]
and therefore it suffices to estimate $\one_{\Delta_{1}}(U)\tilde{R}_{n}\one_{\Delta_{2}}(U)$ for  $\Delta_{i}\Subset \rr^{1+d}$, $i=1,2$.

Writing 
\[
\tilde{R}_{n}=\int \left(\prod_{i=1}^{n}B_{i}^{*}(x_{i})\prod_{i=1}^{n}B_{i}(y_{i})-\prod_{i=1}^{n}B_{i}^{*}(x_{i})B_{i}(y_{i})\right)\prod_{i=1}^{n}\th_{i}(x_{i}, y_{i})d\un xd\un y,
\]
 and commuting $B_{n}^{*}(x_{n})$ to the right, we obtain
 \[
\tilde{R}_{n}= \tilde{R}_{n-1}\circ N_{B_{n}}(\th_{n})+ \sum_{l=1}^{n-1}S_{n,l},
\]
for:
\[
S_{n,l}= \int \left(\prod_{i=1}^{n-1}B_{i}^{*}(x_{i})\prod_{i=1}^{l-1}B_{i}(y_{i})[B_{n}^{*}(x_{n}),  B_{l}(y_{l})]\prod_{i=l+1}^{n}B_{i}(y_{i})\right)\prod_{i=1}^{n}\th_{i}(x_{i}, y_{i})d\un xd\un y,
\]
where  $\prod_{i=1}^{0}B_{i}(y_{i})=1$ is understood.
This implies that:
\beq\bes
\label{eapp.0}
\| \one_{\Delta_{1}}(U)\tilde{R}_{n}\one_{\Delta_{2}}(U)\|_{B(\cH)}   
&\leq  \| \one_{\Delta_{1}}(U)\tilde{R}_{n-1}\one_{\Delta_{3}}(U)\|_{B(\cH)} \| \one_{\Delta_{3}}(U)N_{B_{n}}(\th_{n})\one_{\Delta_{2}}(U)\|_{B(\cH)}  \\   
& \ \ \ + \sum_{l=1}^{n-1}\|  \one_{\Delta_{1}}(U)S_{n,l} \one_{\Delta_{2}}(U)\|_{B(\cH)}\\    
&\leq C(B_{n}) \| \one_{\Delta_{1}}(U)\tilde{R}_{n-1}\one_{\Delta_{3}}(U)\|_{B(\cH)} \|h_{n}\|_{B(L^{2}(\rr^{d}))}\\    
&\ \ \ + \sum_{l=1}^{n-1}\|  \one_{\Delta_{1}}(U)S_{n,l} \one_{\Delta_{2}}(U)\|_{B(\cH)}.
\ees\eeq
The main part of the proof is to estimate $\|  \one_{\Delta_{1}}(U)S_{n,l} \one_{\Delta_{2}}(U)\|_{B(\cH)}$.

Let us fix $u_{i}\in \one_{\Delta_{i}}(U)\cH$ for $\Delta_{i}\Subset \rr^{1+d}$,  $i=1,2$. Then, recalling the definition of 
$a_{\un{B}}=:a_{ B_{1},\ldots, B_{n}}$, we have
 \beq\bes\non
&(u_{1}| S_{n,l}u_{2})_{\cH}\\  
&=\int(\psi_{1}(x_{1}, \dots, x_{n-1})|\prod_{i=1}^{l-1}B_{i}(y_{i})[B_{n}^{*}(x_{n}),  B_{l}(y_{l})]\psi_{2}(y_{l+1}, \dots, y_{n}))_{\cH}\prod_{i=1}^{n}\th_{i}(x_{i}, y_{i})d\un{x}d\un{y},
\ees\eeq
for 
\beq\bes\non
&\psi_{1}(x_{1}, \dots, x_{n-1})= (a_{B_{n-1},\ldots, B_{1}}u_{1})(x_{n-1}, \dots , x_{1}),\\ 
&\psi_{2}(y_{l+1}, \dots, y_{n})= (a_{B_{l+1},\ldots, B_{n}}u_{2})(y_{l+1}, \dots, y_{n}).
\ees\eeq
{\it Step 1:} Let us first perform the integral in  the variables $ x_{1}, \dots, x_{n-1}, y_{n}$. We obtain using (\ref{eapp.-1}):
\beq 
 \bes \label{eapp.1}
(u_{1}| S_{n,l}u_{2})_{\cH}
&=\int(\tilde{\psi}_{1}(y_{1}, \dots, y_{n-1})|\prod_{i=1}^{l-1}B_{i}(y_{i})[B_{n}^{*}(x_{n}), B_{l}(y_{l})]\tilde{\psi}_{2}(y_{l+1}, \dots, y_{n-1}, x_{n}))_{\cH}\times\\
&\quad\times \chi_{l}(y_{l})\chi_{n}(x_{n})dy_{1}\dots dy_{n-1}dx_{n},
\ees
\eeq
for
\beq\bes\non
&\tilde{\psi}_{1}(y_{1}, \dots, y_{n-1})= \big((\th_{1}^{*}\otimes\cdots\otimes\th_{l-1}^{*}\otimes h_{l}^{*}\otimes \th_{l+1}^{*}\otimes \cdots\otimes \th_{n-1}^{*})\psi_{1}\big)(y_{1}, \dots, y_{n-1}),\\
&\tilde{\psi}_{2}(y_{l+1}, \dots, y_{n-1}, x_{n})= \big((\underbrace{\one\otimes\cdots \otimes \one}_{n-l-1}\otimes h_{n})\psi_{2}\big)(y_{l+1}, \dots, y_{n-1}, x_{n}).
\ees\eeq
{\it Step 2:}
We now perform the integrals in $y_{1}, \dots, y_{l-1}$.

Let us first note an easy  fact:  let $v_{1}= v_{1}(y_{1}, \dots, y_{l-1})\in \cH\otimes L^{2}(\rr^{(l-1)d})$ and $v_{2}\in \one_{\Delta}(U)\cH$ for $\Delta\Subset \rr^{1+d}$. Then:
\beq\bes\label{eapp.2}
&\left|\int (v_{1}(y_{1}, \dots, y_{l-1})|\prod_{i=1}^{l-1}B_{i}(y_{i})v_{2})_{\cH}dy_{1}\dots dy_{l-1}\right|\\
&\leq \left(\int \|v_{1}\|^{2}(y_{1}, \dots, y_{l-1}) dy_{1}\dots dy_{l-1}\right)^{\12}\left(\int\| \prod_{i=1}^{l-1}B_{i}(y_{i})v_{2}\|^{2}_{\cH}dy_{1}\dots dy_{l-1}\right)^{\12}\\
&\leq C(B_{1}, \dots, B_{l-1})\|v_{1}\|_{\cH\otimes L^{2}(\rr^{(l-1)d})}\,\, \| v_{2}\|_{\cH},
\ees
\eeq
using successively the Cauchy-Schwarz inequality, the fact that the $B_{i}$ are energy decreasing and Lemma~\ref{existence-lemma}.
Let us denote by $K(y_{1}, \dots, y_{n-1}, x_{n})$ the integrand in (\ref{eapp.1}).  Applying (\ref{eapp.2}) we obtain that:
\beq\bes\non
&\left| \int K( y_{1}, \dots , y_{n-1}, x_{n}) dy_{1}\dots dy_{l-1} \right|\\
&\leq  C(B_{1}, \dots, B_{l-1}) \left(\int \|\tilde{\psi}_{1}(y_{1}, \dots, y_{n-1})\|^{2}_{\cH} dy_{1}\dots dy_{l-1}\right)^{\12}\times\\
&\quad\times \chi_{l}(y_{l})\chi_{n}(x_{n})\| [B_{l}(y_{l}), B_{n}^{*}(x_{n})]\|_{B(\cH)}\| \tilde{\psi}_{2}\|_{\cH}( y_{l+1}, \dots, y_{n-1}, x_{n}).
\ees\eeq
{\it Step 3:}  We perform the remaining integrals in $y_{l}, \dots, y_{n-1}, x_{n}$.

From almost locality we have $\| [B_{l}(y_{l}), B_{n}^{*}(x_{n})]\|_{B(\cH)}\leq C_{N}\langle y_{l}- x_{n}\rangle^{-N}$. Using (\ref{toto.e13bis}), we see that $\chi_{l}(y)\langle y-x\rangle^{-N}\chi_{n}(x)$ is the distributional kernel of $ \chi_{l}g_{N}(D_{x})\chi_{n}$. Using once more the Cauchy-Schwarz inequality we obtain:
\beq\bes\non
&\left|\int K(y_{1}, \dots, y_{n-1}, x_{n}) dy_{1}\dots dy_{n-1}dx_{n}\right|\\
&\leq C(B_{1}, \dots, B_{l-1}) \| \tilde{\psi}_{1}\|_{\cH\otimes L^{2}(\rr^{(n-1)d})}\times \| \chi_{l}g_{N}(D_{x})\chi_{n}\|_{B(L^{2}(\rr^{d}))}\times \|\tilde{\psi}_{2}\|_{\cH\otimes L^{2}(\rr^{(n-l)d})}. 
\ees\eeq

Now since the $B_{i}$, (resp. $\psi_{1}, \psi_{2}$) have compact energy-momentum transfers, (resp. spectrum), we know that:
\[
\|\tilde{\psi}_{1}\|_{\cH\otimes L^{2}(\rr^{(n-1)d})}\leq C(B_{1}, \dots, B_{n-1}) \prod_{i=1}^{n-1}\| h_{i}\|_{B(L^{2}(\rr^{d}))}   \|u_{1}\|_{\cH}, \ \
\| \tilde{\psi}_{2}\|\leq C(B_{n})\| h_{n}\|_{B(L^{2}(\rr^{d}))}   \| u_{2}\|_{\cH}.
\]
Therefore we obtain that
\begin{equation}
\label{eqpp.4}
\|\one_{\Delta_{1}}(U) S_{n, l}\one_{\Delta_{2}}(U)\|_{B(\cH)}\leq C_{N}(B_{1}, \dots , B_{n})\| \chi_{l}g_{N}(D_{x})\chi_{n}\|_{B(L^{2}(\rr^{d}))} \prod_{i=1}^{n}\| h_{i}\|_{B(L^{2}(\rr^{d}))}.
\end{equation}
{\it Step 4:} Making use of  (\ref{eapp.0}) and   (\ref{eqpp.4}) we obtain by induction that
\beqa
\| \one_{\Delta_{1}}(U)\tilde{R}_{n}\one_{\Delta_{2}}(U)\|_{B(\cH)} \leq  C_{N}(\Delta, \un B)\sum_{i\neq j} \| \chi_{i}g_{N}(D_{x})\chi_{j}\|_{B(L^{2}(\real^d))}\prod_{i=1}^{n}\| h_{i}\|_{B(L^{2}(\real^d) ) }. 
\eeqa
(We can start the induction at $n=1$, where the statement is trivial). 
Finally, we estimate the error terms coming from the replacement of $h_{i}$ by $\th_{i}$.  Using that  the operators $a_{B_{1},\ldots, B_{j}}$ are bounded, we obtain that:
 \beq\bes\non
 &\| \one_{\Delta_{1}}(U)(R_{n}- \tilde{R}_{n})\one_{\Delta_{2}}(U)\|_{B(\cH)}\\
&\leq C_{N}(B_{1}, \dots, B_{n}) \times \bigg(   \sum_{i=1}^{n}(\| h_{i}(1- \chi_{i})\|_{B(L^{2}(\rr^{d}))}+ \| (1- \chi_{i})h_{i}\|_{B(L^{2}(\rr^{d}))})\prod_{j\neq i}\| h_{j}\|_{B(L^{2}(\rr^{d}))}\bigg).
\ees\eeq
This completes the proof. \qed

\subsection{Proof of Lemma \ref{tazu}}\label{B-2}

\proof  By Lemma \ref{toto.1} and (\ref{toto.e02}), we know that 
\[
\|N_{B_{1}}(h_{1})\one_{\Delta}(U)\|\leq C \| h_{1}\|_{B(L^{2}(\real^d) )}, \ \|B_{2}^{*}(g_{2})\one_{\Delta}(U)\|\leq C\| g_{2}\|_{L^{2}(\real^d)}.
\]
Therefore, modulo errors controlled by the r.h.s. of (\ref{zorglib}), we can replace $h_{1}$ by $\ti h_{1}= \chi_{1} h_{1}\chi_{1}$ and $g_{2}$ by $\tilde{g}_{2}= \chi_{2}g_{2}$. Arguing as in the proof of \Proposition~\ref{stup} above, we write for $u_{i}\in \one_{\Delta_{i}}(U)\cH$:
\beq\bes\non
& (u_{1}| [N_{B_{1}}(\ti h_{1}), B_{2}^{*}(\ti g_{2})]u_{2})_{\cH}\\
&=\int (u_{1}| [B_{1}^{*}(x_{1})B_{1}(y_{1}), B_{2}^{*}(x_{2})]u_{2})_{\cH}\ti h_{1}(x_{1}, y_{1}) \ti g_{2}(x_{2}) 
dx_{1}dy_{1}dx_{2}\\
&=\int((\one_{\cH}\otimes \ti h_{1}^{*})\circ a_{B_{1}}u_{1}(y_{1})|[B_{1}(y_{1}), B_{2}^{*}(x_{2})]u_{2})_{\cH}\ti g_{2}(x_{2})dy_{1}dx_{2}\\
&\quad+\int( [B_{2}(x_{2}), B_{1}(x_{1})]u_{1}| (\one_{\cH}\otimes \ti h_{1})\circ a_{B_{1}}u_{2}(x_{1}))_{\cH} \ti g_{2}(x_{2})dx_{1}dx_{2}\\
&=I_{1}+ I_{2}.
\ees\eeq
Using Cauchy-Schwarz and almost locality we obtain:
\beq\bes\non
|I_{1}|&\leq C_{N}\int\| (\one_{\cH}\otimes h_{1}^{*}\chi_{1})\circ a_{B_{1}}u_{1}\|_{\cH}(y_{1})\chi_{1}(y_{1})\langle y_{1}- x_{2}\rangle^{-N}\chi_{2}(x_{2})|g_{2}|(x_{2})\| u_{2}\|_{\cH}dy_{1}dx_{2},\\
|I_{2}|&\leq C_{N}\int\| (\one_{\cH}\otimes\chi_{1} h_{1})\circ a_{B_{1}}u_{2}\|_{\cH}(x_{1})\chi_{1}(x_{1})\langle x_{1}- x_{2}\rangle^{-N}\chi_{2}(x_{2})|g_{2}|(x_{2})\| u_{1}\|_{\cH}dy_{1}dx_{2}.
\ees\eeq
Applying once more Lemma \ref{toto.1} we obtain
\[
|I_{1}|+|I_{2}|\leq C_{N}(\Delta_{1}, \Delta_{2})\| h_{1}\|_{B(L^{2}(\real^d))}\|u_{1}\|_{\cH}\|u_{2}\|_{\cH} \| \chi_{1}g_{N}(D_{x})\chi_{2}\|_{B(L^{2}(\real^d))} \|g_{2}\|_{L^{2}(\real^d)},
\]
which completes the proof of the lemma. \qed

\subsection{Proof of Thm.~\ref{Haag-Ruelle}} \label{H-R}

\proof  (1). Let $B_{i},  g_i$ be as specified in the theorem. 
First we show that for $i\neq j$
\begin{equation}
\label{toto.e22}
 [B^{(*)}_{i,t}(g_{i,t}), B^{(*)}_{j,t}(g_{j,t})] \in O(t^{-\infty}).
\end{equation}
 By Prop. \ref{toto.20}
we can find  functions $\chi_{i}, \chi_{j}\in \coinf(\rr^{d})$ with disjoint supports such that
\[
g_{i, t}= \chi_{i}\left(\frac{x}{t}\right)g_{i,t} + O(t^{-\infty})\hbox{ in }
L^{2}(\rr^{d}),
\]
and similarly for $g_{j,t}$. We set $\chi_{i,t}(x):= \chi_{i}(\frac{x}{t})$, 
$\chi_{j,t}(x):= \chi_{j}(\frac{x}{t})$ and note that 
by Lemma \ref{toto.21} (2) 
\[
[B^{(*)}_{i,t}(g_{i,t}), B^{(*)}_{j,t}(g_{j,t})]= [B^{(*)}_{i,t}(\chi_{i,t}g_{i,t}), B^{(*)}_{j,t}(\chi_{j,t}g_{j,t})]+ O(t^{-\infty}).
\]
Using almost locality of $B^{(*)}_{i}, B^{(*)}_{j}$ we obtain from (\ref{quasi-1}) 
that the commutator in the r.h.s. is of order $O(t^{-\infty})$, which proves (\ref{toto.e22}). 
Now  we can write
\begin{align}
 \p_{t}(B_{1,t}^{*}(g_{1,t})\ldots B^*_{n,t}(g_{n,t}))\Omega= \sum_{i=1}^n B_{1,t}^{*}(g_{1,t}) \ldots \pa_t( B_{i,t}^*(g_{i,t}) ) \ldots B^*_{n,t}(g_{n,t})\Omega.   
 \end{align}
Due to Lemma~\ref{toto.21}~(1),  $\pa_t( B_{i,t}^*(g_{i,t}) )$ annihilates the vacuum.
Thus we commute this expression to the right until it acts on the vacuum and show that
 the resulting terms with the commutators are $O(t^{-\infty})$. This follows from  (\ref{toto.e22})   
and from Lemma \ref{toto.21} (2),(3).  Using the Cook argument 
we conclude the proof of  (1). 

Before we proceed to the proof of (2), we need some preparation: 
Let  $B\in\mcL_{0}$ satisfy (\ref{transfer-to-hyperboloid}), and $\Delta= - \supp(\widehat B)\cap \Sp \,U\subset H_{m}$.
We fix $O\subset \rr^{1+d}$, which is an arbitrarily small neighborhood of $\Delta$, and a function $h\in\cS(\rr^{1+d})$ with $\supp\, \widehat{h}\subset O$ and $\widehat{h}= (2\pi)^{-(1+d)/2}$ on $\Delta$. Then $C^{*}:= \int B^{*}(t,x)h(t,x)dtdx$ is an element of $\mcL_{0}$ and
\[
 \widehat{C^{*}}(E, p)= (2\pi)^{(1+d)/2} \widehat{h}(E,p)\widehat{B^{*}}(E,p), \ C^{*}\Omega= (2\pi)^{(1+d)/2} \widehat{h}(H, P)B^{*}\Omega.
\]
Consequently $-\supp(\widehat C)\subset O$, and
\beq\bes\label{toto.e23}
 B^{*}_{t}(g_t)\Omega&= (2\pi)^{d/2}\widehat{f}(P)B^{*}\Om= (2\pi)^{d/2}\widehat{f}(P)\one_{\Delta}(U)B^{*}\Om\\
&=(2\pi)^{d/2}\widehat{f}(P)(2\pi)^{(1+d)/2}\widehat{h}(H, P)B^{*}\Om= (2\pi)^{d/2}\widehat{f}(P)C^{*}\Om= C^{*}_{t}(g_t)\Omega.
\ees
\eeq
We define observables $C_{i}$ corresponding to $B_{i}$ and obtain  
\begin{equation}
\label{toto.e24}
\Psi^{+}= \lim_{t\to\infty} B^{*}_{1,t}(g_{1,t})\ldots B^{*}_{n,t}(g_{n,t})\Om=\lim_{t\to\infty} C^{*}_{1,t}(g_{1,t}) \ldots C^{*}_{n,t}(g_{n,t})\Om,
\end{equation}
where we used (\ref{toto.e22}) and Lemma \ref{toto.21} (2).
Therefore we can assume that the energy-momentum  transfers of $B_{i}^{*}$ entering in the construction of scattering states are localized in arbitrarily small neighborhoods of subsets of $H_{m}$.

(2). Let $\tilde \Psi_t=\tilde{B}^{*}_{1,t}(\tilde g_{1,t})\ldots \tilde{B}^{*}_{n,t}(\tilde g_{n,t})\Om$ be the approximating sequence of the scattering state $\tilde \Psi^+$. In order
to analyse the scalar product $( \tilde \Psi_t|\Psi_t)$ we first show that
\beqa
[[\tilde{B}_{j,t}(\tilde g_{j,t}),B^{*}_{k,t}(g_{k,t})],B^{*}_{l,t}(g_{l,t}) ]\in O(t^{-\infty}), \ k\neq l .\label{double-commutator}
\eeqa
To verify (\ref{double-commutator}) we write  $\tilde g_j=\tilde g_{j,k}+\tilde g_{j,l}$, where $\tilde g_{j,k}$, $\tilde{g}_{j,l}$ are positive energy KG solutions 
such that  $\tilde g_{j,i}$ and  $g_i$  have disjoint velocity supports for $i= k,l$. Then (\ref{double-commutator})
follows from (\ref{toto.e22}) and the Jacobi identity. 

Next we note that
\beqa
\tilde  B_{i,t}(\tilde g_{i,t}) B^{*}_{j,t}(g_{j,t})\Om=\Om( \Om| \tilde B_{i,t}(\tilde g_{i,t})  B^{*}_{j,t}(g_{j,t})\Om) \label{factorization}, \   i,j=1,\ldots, n.
\eeqa
(\ref{factorization}) follows from the fact that  $\tilde  B_{i,t}(\tilde g_{i,t})  B^{*}_{j,t}(g_{j,t})\Om$ belongs 
to the range of $\one_{-K_{j}+\tilde{K}_{i}}(U)$, where $K_j$ and $\tilde K_i$ are the energy-momentum transfers of $B_j$ and $\tilde{B}_i$,
respectively. Due to  (\ref{toto.e24}) $-K_{j}$, $-\tilde{K}_{i}$ can be chosen in arbitrarily small neighbourhoods of $H_m$.
 Since a non-zero vector which is a difference of two vectors from $H_m$ is spacelike, (\ref{factorization}) follows.

\def\tB{{\tilde{B}}}
To prove (\ref{scalar-product}), we set for simplicity of notation $B_{i}(t):= B_{i,t}(g_{i,t})$, $\tilde{B}_{j}(t):= \tilde{B}_{j,t}(\tilde{g}_{j,t} )$. 
We assume that (\ref{scalar-product}) holds for $n-1$ and compute
\beq\bes
\label{toto.e25}
(\tilde{\Psi}_{t}| \Psi_{t})&=(\Om|\tB_n(t)\ldots \tB_1(t) B_1^*(t)\ldots B_n^*(t)\Om)\\ 
&=\sum_{k=1}^n(\Om|\tB_n(t)\ldots \tB_2(t) B_1^*(t)\ldots[\tB_1(t),B_k^*(t)]\ldots B_n^*(t)\Om)\\ 
&=\sum_{k=1}^n\sum_{l=k+1}^n(\Om|\tB_n(t)\ldots \tB_2(t) B_1^*(t)\ldots \check{k} \ldots [[\tB_1(t),B_k^*(t)], B^{*}_l(t)]\ldots B_n^*(t)\Om)\\ 
& \ \ +\sum_{k=1}^n(\Om|\tB_n(t)\ldots \tB_2(t) B_1^*(t)\ldots \check{k} \ldots B_n^*(t)\Om)(\Om| \tB_1(t)B_k^*(t)\Om),
\ees\eeq
where in the last term on the r.h.s. we applied (\ref{factorization}). Now we note that the last term factorizes in the limit $t\to\infty$
by the induction hypothesis and the terms involving double commutators vanish by (\ref{double-commutator}).

It is an immediate consequence of~(\ref{scalar-product}) that the scattering states $\Psi^+$ depend only on the single-particle states $\Psi_i$
(and not on  a particular choice of $B_i$ and $g_i$).  Relation~(\ref{energy-factorization-relation}) follows from
Lemma \ref{toto.21} (1). \qed

\end{document}